\documentstyle[preprint,seceq]{jpsj}

\title{Quantum Effects in Neural Networks}

\author{
Hidetoshi {\sc Nishimori} and 
Yoshihiko {\sc Nonomura}\footnote{Present Address:
Department of Physics, University of Tokyo, 
Hongo, Bunkyo-ku, Tokyo 113, Japan.}
}
 
\inst{
Department of Physics, Tokyo Institute of Technology,\\
Oh-okayama, Meguro-ku, Tokyo 152, Japan
}

\recdate{\hspace{4cm}}

\def\runauthor{Hidetoshi {\sc Nishimori} and Yoshihiko {\sc Nonomura}}
\def\runtitle{Quantum Effects in Neural Networks}

\abst{
We develop the statistical mechanics of the Hopfield model in
a transverse field to investigate how quantum fluctuations
affect the macroscopic behavior of neural networks.
When the number of embedded  patterns is finite, the Trotter
decomposition reduces the problem to that of a random Ising
model.  It turns out that the effects of quantum fluctuations
on macroscopic variables play the same roles as 
those of thermal fluctuations.
For an extensive number of embedded patterns,
we apply the replica method to the Trotter-decomposed system. 
The result is summarized as a ground-state phase diagram 
drawn in terms of the number of patterns per site, $\alpha$, 
and the strength of the transverse field, $\Delta$. 
The phase diagram coincides very accurately with that 
of the conventional classical Hopfield model if we 
replace the temperature $T$ in the latter model by
$\Delta$. Quantum fluctuations are thus concluded to
be quite similar to thermal fluctuations 
in determination of the macroscopic 
behavior of the present model.
}

\kword{
neural networks, Hopfield model, quantum effects, 
macrovariables, phase diagram
}

\begin{document}
\sloppy
\maketitle

%
\section{Introduction}
Statistical mechanics has been applied successfully to 
the analysis of various problems in neural networks. 
In particular, the Hopfield model, a prototype of 
associative memory, was solved explicitly by 
combining well-established techniques
in the mean-field theory of random spin systems. 
In the case of a finite number of embedded patterns,\cite{AGS1}
the application of the mean-field method for the Ising ferromagnet to 
the neural network was shown to prove the existence of a memory-retrieval 
phase as well as mixed (confused) states.  When the number of embedded 
patterns is extensive,\cite{AGS2} it is necessary to introduce 
the replica method in order to investigate the properties of 
the network at finite temperatures, similarly to the case of 
the Sherrington-Kirkpatrick model of spin glasses.\cite{SK}
The resulting phase diagram is characterized by three macroscopic
phases, namely, the retrieval, spin glass and paramagnetic phases.

The introduction of temperature in these studies
was motivated by the apparent
randomness in signal transmission at a synapse:\cite{Amit_book}
A pulse reaching the terminal bulb of an axon does not always
cause the release of neuro-transmitters contained in vesicles.
The probability of release is generally small.\cite{Stapp,Beck_Eccles} 
This randomness in signal transmission
is usually taken into account as thermal fluctuations,
which leads to the stochastic formulation of the problem
in terms of the kinetic Ising model.\cite{Amit_book}
However, detailed considerations of
the origin of this randomness suggest that quantum effects may be
the major driving force to cause uncertainty in the release of
neuro-transmitters from vesicles into the synaptic cleft.
For example, Stapp\cite{Stapp} pointed out that the migration of
calcium ions in the bulb is a quantum diffusion process and thus
the uncertainty in the ion positions leads to quantum fluctuations
in the signal transmission at a synapse. 
Beck and Eccles\cite{Beck_Eccles}
argued that the uncertainty in the positions of hydrogen atoms
in the vesicular grid is the origin of quantum fluctuations of
comparable order of magnitude as thermal fluctuations 
in the brain at room
temperature. These investigations indicate the necessity to treat
randomness in the signal transmission in terms of
quantum mechanics, not simply as thermal fluctuations as
has been the case conventionally.

The relation between quantum mechanics and the brain functioning
has been discussed also in the context to clarify the fundamental
significance
of wave functions and observations in quantum mechanics
(see Refs. \citen{Stapp} and \citen{Penrose} and references therein). 
However, few of the
previous investigations in this area
have paid attention to the behavior of macroscopic
observables. Discussions have been given
mostly in terms of microscopic wave functions, 
though experiments are often carried out on macroscopically
observable variables.  We should point out that the superposition
of various microscopic state vectors does not always lead
to the uncertainty in macroscopic observables.  Quantum fluctuations may
work in a manner similar to thermal fluctuations,
leading only to weak stochastic
deteriorations of observed values of macrovariables.  We will
show this effect explicitly in the present paper.

We therefore have sufficient reasons to introduce quantum fluctuations
into neural networks.  It is in general difficult to reflect directly
microscopic quantum processes in a simple model amenable to analytical 
investigations.   We thus adopt the Hopfield model in a transverse field, 
\begin{equation}
  {\cal H} = -\sum_{(ij)}J_{ij} \sigma_i^z \sigma_j^z
             -\Delta \sum_i \sigma_i^x\ ,
     \label{Hamil_sec1}
\end{equation}
where $\sigma_i^x$ and $\sigma_i^z$ are the components
of a Pauli matrix at site $i$.  The interactions $J_{ij}$ are
given by the Hebb rule as specified explicitly in $\S 2$.
Admittedly this model is not a faithful reproduction of
real processes in the brain.  For example, the state
of a neuron has quantum uncertainty in an eigenstate of the Hamiltonian
(\ref{Hamil_sec1}), quite an improbable situation in reality.
However, our purpose is not to explain the brain itself in detail.
(One may argue in this regard that the Hopfield model without the
transverse-field term is already inadequate as a model of the brain.) 
We rather aim to clarify the role of quantum fluctuations 
in large-scale networks at a phenomenological level.
We believe that the present system (\ref{Hamil_sec1}) 
serves as a first step toward this goal.

This paper is organized as follows.  In $\S 2$ the
transverse-field Hopfield model (\ref{Hamil_sec1}) is solved
for a finite number of embedded patterns.   It is shown that
quantum fluctuations have almost the same effects as thermal
fluctuations on macroscopic properties of the network.  The case with an
extensive number of embedded patters is studied in $\S 3$.
The resulting phase diagram turns out to be almost the same
as that of the classical Hopfield model
qualitatively and even quantitatively. 
(We call the model with $\Delta =0$ in (\ref{Hamil_sec1})
the classical Hopfield model in this paper.)
The last section is devoted to discussions.
A preliminary report of a part of the present work has 
already been given elsewhere.\cite{PRL}
\section{Finite Number of Patterns Embedded}
We first consider the case in which the number of embedded 
patterns $p$ remains finite in the thermodynamic limit.  
\subsection{Formulation}
The Hamiltonian of the Hopfield model in a transverse field 
has already been given in (\ref{Hamil_sec1}) as 
\begin{equation}
  {\cal H} = -\sum_{(ij)}J_{ij} \sigma_i^z \sigma_j^z 
             -\Delta \sum_i \sigma_i^x \equiv {\cal H}_0+{\cal H}_1\ .
     \label{Hamiltonian}
\end{equation}
The interactions $J_{ij}$ obey the Hebb rule, 
\begin{equation}
  J_{ij}=\frac{1}{N}\sum_{\mu =1}^{p}\xi_i^\mu \xi_j^\mu\ ,
  \label{Hebb}
\end{equation}
with $\xi_i^\mu =1$ or $-1$ randomly.  The summation over the 
indices $(ij)$ in (\ref{Hamiltonian}) runs over all combinations
of pairs of sites.  The partition function of this quantum system 
can be represented in terms of simple Ising variables
by the Trotter decomposition,\cite{Suzuki,Usadel}
\begin{equation}
  Z = \lim_{M\to \infty}{\rm Tr}
   \left( {\rm e}^{-\beta {\cal H}_0/M}{\rm e}^{-\beta {\cal H}_1/M}\right) ^M
    = \lim_{M\to \infty} Z_M\ ,
  \label{Trotter}
\end{equation}
where

\begin{full}
\begin{equation}
  Z_M = \sum_{\{ \sigma =\pm 1\} } 
  \exp \left( \frac{\beta}{MN}\sum_{K=1}^M \sum_{(ij)} \sum_{\mu =1}^p
  \xi_i^\mu \xi_j^\mu \sigma_{iK} \sigma_{jK}
    +B \sum_{K=1}^M \sum_{i=1}^N \sigma_{iK} \sigma_{i,K+1} 
    \right)\ .
    \label{ZM}
\end{equation}
\end{full}

\noindent
The coupling constant $B$ in the Trotter direction is related to the
 coefficient $\Delta$ of the transverse field term 
 in (\ref{Hamiltonian}) by\cite{Suzuki,Usadel}
\begin{equation}
  B = \frac{1}{2} \log {\rm cosec} \frac{\beta \Delta}{M}\ .
  \label{B}
\end{equation}
We follow the standard procedure to decompose the double 
summation over $(ij)$ using a Gaussian integral, 

\begin{full}
\begin{equation}
  Z_M =\int \prod_{K\mu} {\rm d}m_{K\mu} \sum_\sigma
   \exp \left( -\frac{N\beta}{2M}\sum_{K\mu} m_{K\mu}^2 
   + \frac{\beta}{M}\sum_{K\mu}\sum_im_{K\mu}\xi_i^\mu \sigma_{iK}
   +B \sum_{Ki} \sigma_{iK} \sigma_{i,K+1} 
    \right)\ ,
   \label{G1}
\end{equation}
\end{full}

\noindent
where we have ignored the overall constant which is irrelevant
for the following arguments.

In the thermodynamic limit $N\to \infty$ with $p$ kept finite,
the saddle point of the integrand of (\ref{G1})
yields the equilibrium free energy per spin as

\begin{full}
\begin{equation}
  f=\frac{1}{2M}\sum_{K\mu}m_{K\mu}^2 -T
  \ll \log \sum_\sigma \exp\left( \frac{\beta}{M}
  \sum_{K\mu}m_{K\mu}\xi^\mu \sigma_K
  +B\sum_K \sigma_K \sigma_{K+1} \right) \gg\ ,
  \label{f1}
\end{equation}
\end{full}

\noindent
where the double brackets $\ll\cdots\gg $ denote the average 
over the randomness of embedded patterns $\{ \xi_i^\mu\}$.
We have assumed the self-averaging property of the
free energy to derive the above expression.\cite{AGS1}
The saddle-point condition leads to the equation of state
\begin{equation}
  m_{K\mu} = \ll \xi^\mu \langle \sigma_K \rangle \gg\ ,
  \label{EOS1}
\end{equation}
where the brackets $\langle \cdots \rangle$ 
stand for the average by the weight
\[
\exp\left( \frac{\beta}{M}
  \sum_{K\mu}m_{K\mu}\xi^\mu \sigma_K
  +B\sum_K \sigma_K \sigma_{K+1} \right)\ .
 \]
Equation (\ref{EOS1}) shows that the parameter $m_{K\mu}$ represents
the overlap of the spin configuration in the $K$th Trotter slice
with the $\mu$th embedded pattern.
The solutions of (\ref{EOS1}) describe the thermodynamic properties
of the system.
%
\subsection{Symmetric solution near the critical point}
Let us first discuss the symmetric solutions of the equation of state
(\ref{EOS1}) in the form $m_{K\mu}=m$ for all $K$ and $\mu\le l$,
$l$ being a given integer.
If $\mu$ exceeds $l$, $m_{K\mu}=0$.  The stability of
this type of solutions
will be considered later. The free energy (\ref{f1}) is now written as

\begin{full}
\begin{equation}
  f= \frac{1}{2} lm^2 
  -T \ll \log \sum_\sigma \exp\left( \frac{\beta m}{M}
  \sum_{K}\sum_{\mu =1}^l \xi^\mu \sigma_K 
  +B\sum_K \sigma_K \sigma_{K+1} \right) \gg\ .
  \label{f2}
\end{equation}
\end{full}

\begin{figure}
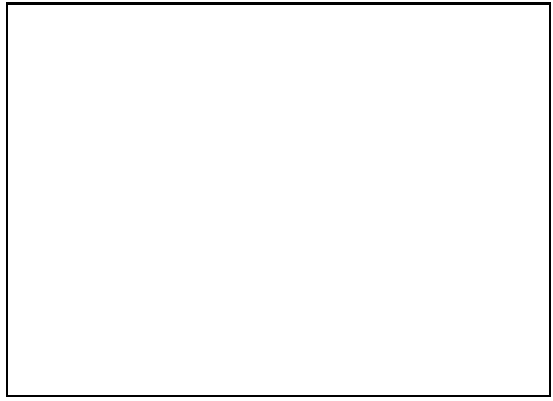

  \figureheight{5cm}
  \caption{The critical line separating the ordered and disordered
  phases.
  }
  \label{fig.1}
\end{figure}

To determine the critical temperature, we expand (\ref{f2}) to $O(m^4)$ as
\begin{eqnarray}
 f&=& \frac{1}{2} lm^2 -T\log Z_0 -\frac{\beta m^2}{2M^2}
     \ll z_l^2 \gg \langle \left( \sum_K \sigma_K \right) ^2
     \rangle_{\rm 0c}
     \nonumber \\
     &-&\frac{\beta^3 m^4}{24M^4}  \ll z_l^4 \gg
     \langle \left( \sum_K \sigma_K \right) ^4
     \rangle_{\rm 0c}\ ,
   \label{f3}
\end{eqnarray}
where
\[
   Z_0 = \sum_\sigma {\rm e}^{B\sum_K \sigma_{K} \sigma_{K+1}}
\]
and
\[
   z_l = \sum_{\mu =1}^{l} \xi^\mu\ .
\]
In (\ref{f3}) the brackets $\langle\cdots\rangle_{\rm 0c}$
represent the cumulants calculated from the average
$\langle\cdots\rangle_0$ defined by 
\begin{equation}
  \langle Q\rangle_{0} \equiv \frac{1}{Z_0}\sum_\sigma
  Q \, {\rm e}^{B\sum_K \sigma_{K} \sigma_{K+1}}\ .
  \label{Ave0}
 \end{equation}
In order to evaluate the cumulants appearing in (\ref{f3}),
we calculate the generating function
\begin{equation}
  Z_{h,M} = \sum_\sigma \exp \left( B\sum_K \sigma_{K} \sigma_{K+1}
  +\frac{h}{M}\sum_K \sigma_K \right)\ .
\label{ZhM}
\end{equation}
In the limit of large $M$, we find,
using the Trotter decomposition formula:
\begin{eqnarray}
  Z_h &\equiv & \lim_{M\to\infty} Z_{h,M} = {\rm Tr}\, {\rm e}^{h\sigma_z
   +\beta \Delta \sigma_x }
   \nonumber \\
   &=& 2\cosh \sqrt{h^2+\beta^2 \Delta^2}\ .
  \label{single-Z}
\end{eqnarray}
Expansion of this equation in powers of $h$ gives
\begin{eqnarray}
  \log Z_h&=&\log Z_0 \Bigr|_{M\to\infty}+\frac{h^2\tanh a}{2a}\nonumber\\
          & &+\frac{h^4(a-\tanh a-a\tanh^2 a)}{12a^3}\ ,
  \label{Zh2}
\end{eqnarray}
%
with $a=\beta \Delta$. Comparison of (\ref{Zh2}) with (\ref{f3})
gives the cumulants in the limit $M\to\infty$ as
\begin{eqnarray}
 \frac{1}{M^2}\langle \left( \sum_K \sigma_K \right) ^2
     \rangle_{\rm 0c} &=& \frac{\tanh{a}}{a}\ ,
   \nonumber \\
  \frac{1}{M^4} \langle \left( \sum_K \sigma_K \right) ^4
     \rangle_{\rm 0c}&=&\frac{3}{a^3}(a-\tanh a-a\tanh^2a)\ .
  \nonumber
\end{eqnarray}
Using the relations
\begin{eqnarray}
  \ll z_l^2 \gg &=& l \label{zl2}\ ,\\
   \ll z_l^4 \gg &=& l(3l-2)\ ,
   \label{zl4}
\end{eqnarray}
we finally obtain the explicit expression of the free energy (\ref{f3}) as
\begin{eqnarray}
  && \frac{f}{l} =\frac{1}{2}m^2 
        -\frac{\beta m^2}{2}\cdot \frac{\tanh a}{a}
        \nonumber \\
    &&-(3l-2)\cdot\frac{\beta^3 m^4}{8}
    \cdot\frac{a-\tanh a-a\tanh^2 a}{a^3}\ .
  \label{f4}
\end{eqnarray}
The critical condition is thus expressed as
\[
   \beta_{\rm c} \frac{\tanh a}{a} =1
 \]
or
\begin{equation}
  \tanh \beta_{\rm c} \Delta = \Delta\ .
\label{Tc1}
\end{equation}
This formula shows that the critical temperature does not depend on
the parameter $l$ similarly to the case of the classical model.\cite{AGS1}
The critical line (\ref{Tc1}) is drawn in Fig. \ref{fig.1}.
The asymptotic form of the overlap order parameter around the
critical temperature is derived from (\ref{f4}) as
\begin{equation}
  m^2 = \frac{2\left( 1-\Delta^{-1}\tanh \beta \Delta \right)}
             {\beta_c^3 (3l-2)g(a_{\rm c})}\ ,
 \label{m1}
\end{equation}
where $a_{\rm c}=\beta_{\rm c} \Delta$ and
\[
  g(a) = \frac{1}{a^3}(a-\tanh a-a\tanh^2 a)\ .
 \]
%
%
\subsection{Symmetric solution in the ground state}
In the limit of large $M$,
the free energy (\ref{f2}) is written as
\[
 f = \frac{1}{2}lm^2 -T\ll \log {\rm Tr} \, {\rm e}^{\beta mz_l\sigma_z
  +\beta\Delta \sigma_x } \gg
\]
according to the Trotter decomposition formula.  This expression
is further evaluated in the limit $T\to 0$ as
\begin{eqnarray}
  f&=&\frac{1}{2}lm^2 -T\ll \log  2\cosh \beta\sqrt{m^2 z_l^2
        +\Delta^2} \gg
        \nonumber \\
   &\longrightarrow &\frac{1}{2}lm^2 - \ll \sqrt{m^2 z_l^2
        +\Delta^2} \gg\ .
    \label{f5}
\end{eqnarray}
The equation of state is obtained from (\ref{f5}) as
\begin{equation}
   lm=\ll \frac{mz_l^2}{\sqrt{m^2 z_l^2+\Delta^2}} \gg\ .
  \label{EOS2}
\end{equation}
For a finite value of the overlap order parameter $m$, 
(\ref{EOS2}) reads
\[
     l=\ll \frac{z_l^2}{\sqrt{m^2 z_l^2+\Delta^2}} \gg\ . 
  \]
In consideration of (\ref{zl2}), this equation implies that $m$ 
decreases to 0 as $\Delta$ approaches 1. Thus the critical value of 
$\Delta$ in the ground state is 1.  The free energy (or the energy) at 
this critical point is $-1$ independent of $l$ as is apparent from (\ref{f5}).

\begin{figure}
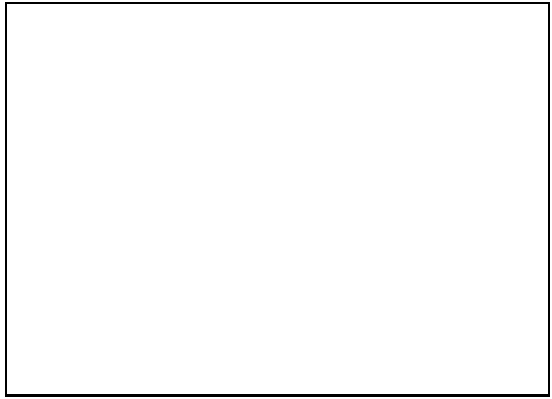

  \figureheight{5cm}
  \caption{The ground-state magnetization as a function
  of $\Delta$ for $l=1, 2$ and 3.}
  \label{fig.2}
\end{figure}

\begin{figure}
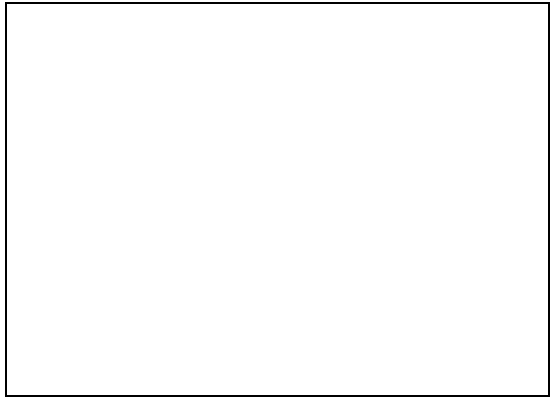

  \figureheight{5cm}
  \caption{The ground-state free energy (or the energy) as a function
  of $\Delta$ for $l=1, 2$ and 3.
  }
  \label{fig.3}
\end{figure}

It is straightforward to derive the explicit forms of the order 
parameter and the free energy for small values of $l$ from
(\ref{EOS2}) and (\ref{f5}). The results are given by 
\begin{equation}
  m=\sqrt{1-\Delta^2},~~~f=-\frac{1}{2}(1+\Delta^2 )
  \label{l1}
\end{equation}
and
\begin{equation}
  m=\frac{1}{2}\sqrt{1-\Delta^2},~~~f=-\frac{1}{4}(1+\Delta)^2
  \label{l2} 
\end{equation}
for $l=1$ and $l=2$, respectively, and
\begin{eqnarray}
  & &\frac{3}{\sqrt{9m^2 +\Delta^2}}+\frac{1}{\sqrt{m^2 +\Delta^2}}=4\ ,
    \nonumber \\
  f&=&\frac{3}{2}m^2 -\frac{1}{4}\left( \sqrt{9m^2 +\Delta^2}
    +3\sqrt{m^2 +\Delta^2} \right)
  \label{l3}
\end{eqnarray}
for $l=3$.  These results are plotted in Fig.~\ref{fig.2} for the 
order parameter and in Fig.~\ref{fig.3} for the free energy.
These figures show that in the ground state quantum effects represented by
the parameter $\Delta$  play  very similar roles to
thermal fluctuations represented by the temperature $T$
in the classical model\cite{AGS1} in which
the order parameter and the free energy behave in
almost the same way as functions of $T$ as in
Figs. \ref{fig.2} and \ref{fig.3}
if we replace $\Delta$ by $T$.
%
%
\subsection{Stability of the symmetric solutions
 {\rm (I)}}
It is necessary to check the stability of the symmetric 
solutions given in the previous subsection. The overlap
order parameter $m_{K\mu}$ should depend upon $K$ and
$\mu$ in general.  The possible dependence of $m_{K\mu}$ 
on the Trotter number $K$ is
investigated in the present subsection.  The pattern number
dependence will be treated in the next subsection.

Let us assume that $m_{K\mu}=m_K$ if $\mu\le l$ and $m_{K\mu}=0$
otherwise. Then, the free energy (\ref{f1}) is expanded to
second order of $m_K$ as

\begin{full}
\begin{equation}
  f = -T\log Z_0 +\frac{l}{2M}\sum_K m_K^2 -\frac{\beta l}{2M^2}
       \sum_{KK'} m_K m_{K'} \langle \sigma_K \sigma_{K'} \rangle_0
    =-T\log Z_0 +\frac{l}{2M}\sum_{KK'} A_{KK'}m_K m_{K'}\ ,
      \label{f6}
\end{equation}
\end{full}

\noindent
where
\begin{equation}
  A_{KK'} = \delta_{KK'} -\frac{\beta}{M}
      \langle \sigma_K \sigma_{K'} \rangle_0\ .
  \label{AKK}
\end{equation}
The average $\langle\cdots\rangle_0$ was defined
in (\ref{Ave0}).
The expansion of the free energy (\ref{f6})  corresponds to the approach
to the critical point from the paramagnetic
phase because the order parameter vanishes in equilibrium in the
high-temperature paramagnetic phase.  Correspondingly, from (\ref{AKK}),
all eigenvalues of the matrix $A_{KK'}$
are seen to be positive for small $\beta$ or large $T$,
implying the stability of the paramagnetic solution $m_K=0$. 
The critical temperature is determined by the
vanishing point of the minimum eigenvalue $\lambda_0$
of the matrix $A_{KK'}$,
\begin{eqnarray}
   \lambda_0 &=&1-\frac{\beta}{M^2} \sum_{KK'}
      \langle \sigma_K \sigma_{K'} \rangle_0 
      \nonumber \\
    &=& 1-\frac{\beta}{M^2}
        \langle \left( \sum_K \sigma_K \right) ^2\rangle_0 
     \nonumber \\
    &
       \mathop{\longrightarrow}_{M\to \infty}
       & 1-\frac{\tanh \beta \Delta}{\Delta}\ .
   \label{lambda_0}
\end{eqnarray}
The critical condition $\tanh \beta_{\rm c} \Delta =\Delta$
of (\ref{Tc1}) is thus recovered.  An important point here is that 
the eigenstate corresponding to the minimum eigenvalue $\lambda_0$
is a uniform mode $m_K =m$. This means that an ordered state 
uniform in the Trotter number $K$
is formed slightly below the critical temperature. Therefore, 
we conclude that the symmetric solution $m_{K\mu}=m$ is stable
against the Trotter-number dependence slightly below the critical
temperature.

At an arbitrary temperature the free energy is written as

\begin{full}
\begin{equation}
  f=\frac{l}{2M}\sum_{K}m_{K}^2 
  -T \ll \log \sum_\sigma \exp\left( \frac{\beta z_l}{M}
  \sum_{K}m_{K} \sigma_K 
  +B\sum_K \sigma_K \sigma_{K+1} \right) \gg\ .
  \label{f7}
\end{equation}
\end{full}

\noindent
To investigate the stability of the symmetric solution, we check the
eigenvalues of the Hessian

\begin{full}
\begin{equation}
    \left. A_{KL}=\frac{\partial^2 f}{\partial m_K \partial m_L}
    \right|_{m_K=m}
    = \frac{l}{M}\delta_{KL} 
    -\frac{\beta}{M^2}
      \ll z_l^2 \left( \langle \sigma_K \sigma_L \rangle
          -\langle \sigma_K \rangle\langle \sigma_L \rangle
           \right) \gg\ ,
      \label{Hess}
\end{equation}
\end{full}

\noindent
where the brackets $\langle \cdots \rangle $ denote the average 
with respect to the weight
%
\[
  \exp\left( \frac{\beta z_l m}{M}
  \sum_{K} \sigma_K
  +B\sum_K \sigma_K \sigma_{K+1} \right)\ .
\]
Since the value of the matrix element $A_{KL}$ depends only on 
the difference $|K-L|$ and the quantity in the double brackets 
$\ll\cdots\gg$ in (\ref{Hess}) is positive, the lowest 
eigenvalue of this Hessian is given by 
\[
  \lambda_0 = \sum_{L=1}^{M} A_{KL}\ .
\]
The symmetric solution is stable if $\lambda_0$ is positive. Explicitly, 

\begin{full}
\begin{eqnarray}
 M\lambda_0 &= &l-\beta \frac{\partial^2}{\partial (\beta m)^2}
            \ll \log \sum_\sigma \exp\left( \frac{\beta z_lm}{M}
            \sum_{K}\sigma_K
           +B\sum_K \sigma_K \sigma_{K+1} \right) \gg 
        \nonumber \\
      & \mathop{\longrightarrow}_{M\to \infty} &
     l - T\frac{\partial ^2}{\partial m^2}
            \ll \log {\rm Tr} \, {\rm e}^{\beta z_l m\sigma_z 
            +\beta\Delta\sigma_x}\gg
         \nonumber \\
   &  =&l-\Delta ^2 \ll \frac{z_l^2}{(m^2 z_l^2 +\Delta^2 )^{3/2}}
          \tanh \beta\sqrt{m^2 z_l^2 +\Delta^2} \gg
           \nonumber \\
   & &   \hspace{2cm}-\beta m^2 \ll \frac{z_l^4}{m^2 z_l^2 +\Delta^2}
          {\rm sech}^2 \beta\sqrt{m^2 z_l^2 +\Delta^2} \gg\ .
  \label{lambda_1}
\end{eqnarray}
\end{full}

\noindent
At $T=0$, the above expression reduces to
\[
   M\lambda_0 =l-\Delta^2 \ll \frac{z_l^2}{(m^2 z_l^2 
   +\Delta^2 )^{3/2}} \gg\ .
 \]
For $\Delta\to 0$, $M\lambda_0$ tends to $l$, a positive 
value. If, on the other hand, $\Delta$ is close to 1
($\Delta =1-\epsilon$ with $\epsilon\ll 1$),
$M\lambda_0$ approaches $2\epsilon l$, again positive. We thus 
expect that the eigenvalue $\lambda_0$ remains positive 
between these two limiting values of $\Delta$ when $T=0$.

In the case of general finite temperature,
we have numerically confirmed the positivity of $\lambda_0$ for
$l=3$.  It is expected that the same property holds for other 
values of $l$, though it is difficult to prove it explicitly.
%
\subsection{Stability of the symmetric solution {\rm (II)}}
We next check the pattern number ($\mu$) dependence of the solution
of the equation of state.  If we assume $m_{K\mu}=m_\mu$ for
$\mu\le l$ and $m_{K\mu}=0$ otherwise, the free energy (\ref{f1}) reads

\begin{full}
\begin{eqnarray}
  f&=& \frac{1}{2}\sum_{\mu}m_\mu^2-T\ll \log\sum_\sigma \exp \left(
     \frac{\beta}{M}\sum_{\mu}m_\mu\xi^\mu\sum_K\sigma_K
     +B\sum_K \sigma_K \sigma_{K+1}\right) \gg
   \nonumber \\
   &
        \mathop{\longrightarrow}_{M\to \infty}
    &\frac{1}{2}{\mib m}^2 -T\ll \log 2\cosh \beta\sqrt{({\mib m}\cdot
      {\mib \xi })^2+\Delta^2} \gg\ .
    \label{f8}
\end{eqnarray}
\end{full}

\noindent
The vector ${\mib m}$ has components
$(m_1, m_2,\cdots,m_l,0,0,\cdots )$, and similarly for ${\mib \xi}$.
The Hessian around the symmetric solution is given by

\begin{full}
\begin{eqnarray}
A_{\mu\nu }& =& \left.\frac{\partial^2 f}{\partial m_\mu \partial m_\nu}
     \right| _{m_\mu =m} 
    \nonumber \\
   & =& \delta_{\mu\nu}-\Delta^2 
    \ll \frac{\xi^\mu\xi^\nu}{(m^2 z_l^2+\Delta^2)^{3/2}}
             \tanh \beta\sqrt{m^2 z_l^2+\Delta^2} \gg
    \nonumber \\
  & &\hspace{1cm}-\beta m^2\ll \frac{z_l^2 \xi^\mu\xi^\nu}{m^2 z_l^2+\Delta^2}
       {\rm sech}^2 \beta\sqrt{m^2 z_l^2+\Delta^2} \gg\ .
     \label{Hess2}
\end{eqnarray}
\end{full}

\noindent
The diagonal element of $\{ A_{\mu\nu}\} $
is written in the following form
\begin{eqnarray}
     e_1\equiv A_{\mu\mu}&=&1-\Delta^2 \ll 
      \frac{\tanh \beta\sqrt{m^2 z_l^2+\Delta^2}}
        {(m^2 z_l^2+\Delta^2)^{3/2}} \gg
      \nonumber \\
    &&+\beta q-\beta m^2\ll\frac{z_l^2}{m^2 z_l^2+\Delta^2} \gg\ ,
    \label{diag}
\end{eqnarray}
where
\begin{eqnarray}
  q &=& \ll\frac{m^2 z_l^2}{m^2 z_l^2+\Delta^2} 
  \tanh^2 \beta \sqrt{m^2 z_l^2+\Delta^2}
  \gg
     \nonumber \\
   &=& \ll \langle \sigma_K \rangle^2 \gg\ .
   \label{q}
\end{eqnarray}
All off-diagonal elements have the same value:
\begin{eqnarray}
  &&e_2 \equiv  A_{\mu\nu} ~~~(\mu\ne\nu)
  \nonumber\\
  &&=-\Delta ^2 \ll \frac{\xi^\mu \xi^\nu}
      {(m^2 z_l^2+\Delta^2)^{3/2}} 
            \tanh \beta\sqrt{m^2 z_l^2+\Delta^2} \gg
          \nonumber\\
  & &-\beta m^2 \ll\frac{z_l^2\xi^\mu \xi^\nu}{m^2 z_l^2+\Delta^2}
      {\rm sech}^2 \beta\sqrt{m^2 z_l^2+\Delta^2} \gg\ .
      \label{off-diag}
\end{eqnarray}
{}From (\ref{Hess2}), (\ref{diag}) and (\ref{off-diag}),
it is easy to see that there are three eigenvalues
of the Hessian matrix $\{ A_{\mu\nu}\}$,
\begin{subeqnarray}
  & &\lambda_1 =e_1 +(l-1)e_2\ , \label{le1}\\
  & &\lambda_2 =e_2\ ,  \label{le2}\\
  & &\lambda_3 =e_1 -e_2\ , \label{le3}
\end{subeqnarray}
with the degeneracy $p-l, 1$ and $l-1$, respectively.

To investigate the stability of the symmetric solution near
the critical temperature, we expand various terms in (\ref{diag})
and (\ref{off-diag}) to second order of $m$ using (\ref{q}):
\begin{eqnarray}
  & &m^2 \ll \frac{z_l^2}{m^2 z_l^2+\Delta^2 } \gg \approx 
        \frac{m^2 l}{\Delta}\ ,
        \nonumber\\  
   & & q\approx  m^2 l\cdot\frac{\tanh^2\beta\Delta }{\Delta^2} 
      \nonumber
\end{eqnarray}
and

\begin{full}
\begin{equation}
 \ll\frac{\tanh \beta\sqrt{m^2 z_l^2+\Delta^2}}
            {(m^2 z_l^2+\Delta^2)^{3/2}} \gg
            \approx  \frac{1}{\Delta^3}
            \left(\tanh\beta\Delta \right.
            \nonumber\\
            -\frac{3m^2l}{2\Delta^2}\tanh\beta\Delta
            +\left. \frac{\beta m^2 l}{2\Delta\cosh^2\beta\Delta }
            \right)\ .
\end{equation}
\end{full}

\noindent
These equations and (\ref{m1}) yield
\begin{eqnarray}
  e_1&\approx & \frac{2}{3l-2}
         \left(\frac{\tanh\beta\Delta}{\Delta}-1\right)
            \approx \frac{2(1-\Delta^2)}{3l-2}\epsilon\ ,
     \nonumber\\
  e_2 &\approx & \frac{6\left(\Delta^{-1}\tanh\beta\Delta-1\right)}
      {3l-2} \approx \frac{3(1-\Delta^2)}{3l-2}\epsilon\ ,
     \nonumber
\end{eqnarray}
where $\epsilon $ represents the deviation 
of the temperature from the critical value, 
\[ \epsilon = T_{\rm c}(\Delta )-T\ .\]
The signs of the eigenvalues in (\ref{le1})
are then found as $\lambda_1, \lambda_2 >0$ and
$\lambda_3 <0$.
Since the degeneracy of the last eigenvalue $\lambda_3$
is $l-1$, the symmetric solution with $l\ge 2$ is unstable against
fluctuations in the direction of the eigenvector corresponding to
$\lambda_3$ slightly below the critical temperature.  This
property is exactly the same as in the case of the classical
Hopfield model.\cite{AGS1}

When $T=0$, the matrix elements (\ref{diag}) and (\ref{off-diag})
reduce to
\begin{eqnarray}
     e_1 &=& 1-\Delta^2 \ll\frac{1}{(m^2 z_l^2+\Delta^2)^{3/2}} \gg\ ,
       \nonumber\\
     e_2 &=& -\Delta^2 \ll \frac{\xi^\mu\xi^\nu}
           {(m^2 z_l^2+\Delta^2)^{3/2}} \gg\ .
        \nonumber
\end{eqnarray}
For $l=1$, $\lambda_1=e_1=1-\Delta^2\ge 0$,
implying the stability as expected trivially for the one-component 
solution.  When $l=2$, the eigenvalues are given by 
\begin{subeqnarray}
  \lambda_1&=&1-\Delta^2\ ,
        \label{le4} \\
  \lambda_2&=&1-\frac{\Delta^2}{2}-\frac{1}{2\Delta}\ ,
        \label{le5} \\
   \lambda_3&=& 1-\frac{1}{\Delta}\ .
        \label{le6}
\end{subeqnarray}
The first eigenvalue $\lambda_1$ is always positive in the range
$0\le\Delta <1$, while the second one $\lambda_2$ is positive for 
$\Delta >0.618$ and is negative for $\Delta <0.618$. The third 
eigenvalue $\lambda_3$ is always negative. Thus, this $l=2$ 
solution is unstable at $T=0$ for any $\Delta$ between 0 and 1. 

\begin{figure}
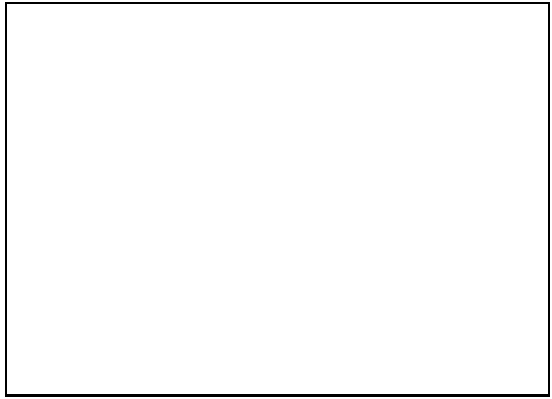

  \figureheight{5cm}
  \caption{Three eigenvalues of the Hessian for $l=3$ and $T=0$.
  The third eigenvalue $\lambda_3$ crosses 0 at $\Delta =0.494$.
  }
  \label{fig.F2}
\end{figure}

\begin{figure}
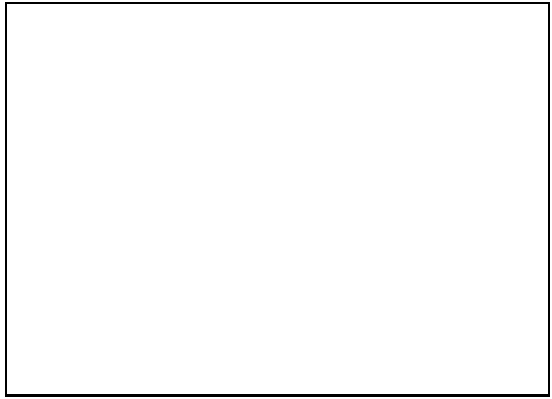

  \figureheight{5cm}
  \caption{Three Hessian eigenvalues for $l=4$ and $T=0$.
  }
  \label{fig.F3}
\end{figure}

The symmetric solution with $l=3$ has the following eigenvalues
for $T=0$:
\begin{eqnarray}
 &&\lambda_1=1-\frac{\Delta^2}{4}
   \left[\frac{3}{(9m^2+\Delta^2)^{3/2}}+\frac{1}{(m^2+\Delta^2)^{3/2}}
        \right]\ ,
        \nonumber \\
  &&\lambda_2=1-\frac{\Delta^2}{4}
   \left[\frac{1}{(9m^2+\Delta^2)^{3/2}}+\frac{3}{(m^2+\Delta^2)^{3/2}}
        \right]\ ,
        \nonumber \\
   &&\lambda_3= 1-\frac{\Delta^2}{(m^2+\Delta^2)^{3/2}}\ .
        \nonumber
\end{eqnarray}
These eigenvalues are plotted as functions of $\Delta$ in Fig.~\ref{fig.F2}.
The first and second eigenvalues $\lambda_1$ and $\lambda_2$
are both positive, but $\lambda_3$ changes its sign at $\Delta =0.494$.
Similarly, the behavior of the eigenvalues for $l=4$ and $l=5$ is shown
in Figs.~\ref{fig.F3} and \ref{fig.F4}. 
\begin{figure}
  \figureheight{5cm}
  \caption{Three eigenvalues of the Hessian for $l=5$ and $T=0$.
  The third eigenvalue $\lambda_3$ crosses 0 at $\Delta =0.426$.
  }
  \label{fig.F4}
\end{figure}
This analysis for $l$ up to 5 thus suggests that the even-$l$ solution
is always unstable while the odd-$l$ solution is stable in a finite
range $0\le \Delta <\Delta_{\rm c}$ when $T=0$.

To confirm this conjecture, we have expanded the eigenvalues
$\lambda_1, \lambda_2$ and $\lambda_3$ as functions of
$\epsilon =1-\Delta$ for small $\epsilon$.  
The equation of state (\ref{EOS2}) has the solution
$m=c\sqrt{\epsilon}$ with $c=\sqrt{2/(3l-1)}$. This gives 
\begin{eqnarray}
  e_{1}&=&\frac{1}{2}(3lc^2-2)\epsilon\ ,
   \nonumber\\
  e_{2}&=&3c^2\epsilon\ . 
   \nonumber
\end{eqnarray}
%
Then, the eigenvalues (\ref{le1}) are 
\begin{subeqnarray}
    \lambda_1 &=& 2\epsilon\ , \nonumber\\
    \lambda_2 &=& \frac{2\epsilon}{3l-2}\ , \nonumber\\
    \lambda_3 &=& -\frac{3l+2}{3l-2}\epsilon\ .\nonumber
\end{subeqnarray}
Since the third eigenvalue $\lambda_3$
has degeneracy $l-1$ and is negative,
we conclude that all symmetric solutions with $l\ge 2$ are
unstable near $\Delta =1$.  This, together with the analysis of
eigenvalues for $l=2$ to 5 explained before, suggests that
the coefficient $\Delta$ of the transverse-field term
 of the present model in the
ground state has effects very similar to
the temperature of the classical Hopfield model
($\Delta =0$) in which the
even-$l$ solutions are unstable for any $T$ while the
odd-$l$ solutions are stable below certain temperatures.\cite{AGS1}

\begin{figure}
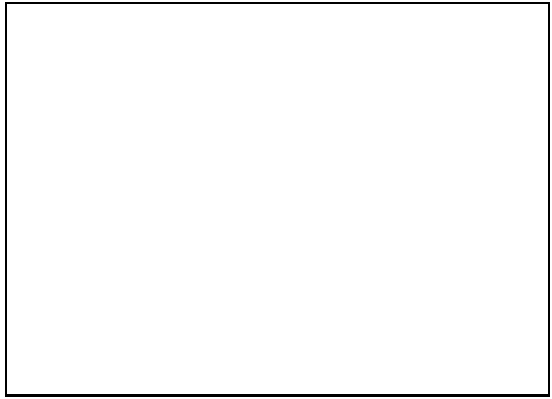

  \figureheight{5cm}
  \caption{
  An asymmetric solution of the type
  $(m,m,u,u,u,0,0,\cdots )$ for $T=0$.
  }
  \label{fig.F5}
\end{figure}

The instability of the even-$l$ solutions suggests the existence of
asymmetric solutions of the equation of state (\ref{EOS1})
as was the case in the classical model.\cite{AGS1}
We actually have found many asymmetric solutions.
An example of a solution of the type
$(m,m,u,u,u,0,0,\cdots )$
is shown in Fig.~\ref{fig.F5}.
Although we were not able to find general rules for
the existence and behavior of asymmetric solutions,
we observed that, when $T=0$, $\Delta$ has effects very similar to
$T$ in the classical model.

\section{Extensive Number of Patterns Embedded}
In this section the number of patterns embedded, $p$,
is assumed to be proportional to the system size $N$.
We closely follow Chap. 10 of Ref. \citen{Hertz}
in our presentation.
%
\subsection{General form of the free energy}
According to (\ref{G1}), 
the replicated partition function averaged over the quenched
randomness is written for a finite Trotter number $M$ as

\begin{full}
\begin{eqnarray}
  \ll Z^n \gg
 &=&\ll \int \prod_{K\mu\rho} {\rm d}m_{K\mu\rho} \sum_\sigma
   \exp \left( -\frac{N\beta}{2M}\sum_{K\mu\rho} m_{K\mu\rho}^2 
   \right.
     \nonumber\\
 &&\hspace{3cm}  \left.
 +\frac{\beta}{M}\sum_{K\mu\rho}\sum_im_{K\mu\rho}\xi_i^\mu 
   \sigma_{i\rho K}
   +B \sum_{Ki\rho} \sigma_{i\rho K} \sigma_{i\rho,K+1} 
    \right) \gg\ ,
   \label{Zn}
\end{eqnarray}
\end{full}

\noindent
where $\rho\ (=1,\cdots,n)$ represents the replica index.
The overall constant is irrelevant and is ignored here. 

We consider the situation in which a finite number of patterns 
have nonvanishinig overlaps.  That is,  $m_{K\mu\rho}$ is 
of order unity for $\mu =1,2,\cdots ,s$ and is of order 
$1/\sqrt{N}$ for other $\mu$'s with $s$ being a finite number. 
Then, the configurational average $\ll\cdots\gg$ in (\ref{Zn}) 
can be evaluated explicitly for patterns $\mu >s$. 
For the $\mu$th pattern with $\mu >s$,

\begin{full}
\begin{eqnarray}
  \prod_i \ll \exp \left( \frac{\beta}{M}\xi_i^\mu
      \sum_{K\rho } m_{K\mu\rho}  \sigma_{i\rho K} \right) \gg
  &=&\prod_i\cosh \left( \frac{\beta}{M}
     \sum_{K\rho } m_{K\mu\rho}  \sigma_{i\rho K} \right) 
     \nonumber \\
  &\approx&\exp \left( \frac{\beta^2}{2M^2}\sum_i\sum_{\rho\sigma}
            \sum_{KL}m_{K\mu\rho} m_{L\mu\sigma} 
                 \sigma_{i\rho K} \sigma_{i\sigma L}  \right)\ .
            \label{e1}
\end{eqnarray}
\end{full}

\noindent
The terms containing $\{ m_{K\mu\rho}\}_{\mu >s}$ in (\ref{Zn})
and (\ref{e1}) are collected in a single formula, 
%
\[
  E_\mu \equiv \exp\left( -\frac{\beta N}{2M}\sum_{K\rho}\sum_{L\sigma}
      \tilde{\Lambda}_{K\rho ,L\sigma}m_{K\mu\rho}m_{L\mu\sigma}
            \right)\ ,
\]
where
\begin{equation}
    \tilde{\Lambda}_{K\rho ,L\sigma}=\delta_{K\rho ,L\sigma}
          -\frac{\beta}{NM}\sum_i \sigma_{i\rho K} \sigma_{i\sigma L}\ .
          \label{Lambda-tilde}
\end{equation}
The integration of $E_\mu$ over $\{ m_{K\mu\rho}\}$ gives
\[
    \int\prod_{K\rho}{\rm d}m_{K\mu\rho}E_\mu
      = {\rm const.}\times(\det \tilde{\Lambda})^{-1/2}\ .
\]
The product of this result over $\mu =s+1,\cdots,p$ is
\begin{eqnarray}
  (\det \tilde{\Lambda})^{-(p-s)/2} 
  &\approx&(\det \tilde{\Lambda})^{-p/2}\nonumber\\
        &=&\exp\left( -\frac{p}{2}\sum_{K\rho}\log {\tilde \lambda}_{K\rho}
               \right)\ .\nonumber
\end{eqnarray}
Here we have neglected the finite number $s$ in the exponent
since it is small compared to the extensive number $p$.  
The eigenvalues of the matrix $\tilde \Lambda$ are
denoted as ${\tilde \lambda}_{K\rho}$.

The eigenvalue ${\tilde \lambda}_{K\rho}$ depends on the
spin configuration as is apparent from (\ref{Lambda-tilde}).
To avoid this complication, we introduce another matrix $\Lambda$
without explicit dependence upon the spin configurations:
\begin{equation}
  \Lambda_{K\rho ,L\sigma}=\delta_{K\rho ,L\sigma}
          -\frac{\beta}{M}q_{\rho\sigma}(KL)-\delta_{\rho\sigma}
            \frac{\beta}{M}S_\rho (KL)\ .
     \label{Lamb}
\end{equation}
This matrix $\Lambda$ is equal to $\tilde{\Lambda}$, if
\begin{equation}
    q_{\rho\sigma}(KL) =\left\{ 
      \begin{array}{ll}
        {\displaystyle \frac{1}{N}\sum_i\sigma_{i\rho K}\sigma_{i\sigma L}}
                       &(\rho\ne\sigma)
            \\
          0 & (\rho =\sigma )
    \end{array}
    \right.
  \label{qKL}
\end{equation}
and
\begin{equation}
    S_\rho (KL)= \frac{1}{N}\sum_i\sigma_{i\rho K}\sigma_{i\rho L}
    \label{SKL}
\end{equation}
are satisfied. 
Physically, $ q_{\rho\sigma}(KL)$ is the spin glass order parameter 
for spins at the Trotter slices $K$ and $L$, and $S_\rho (KL)$ 
is a measure of quantum fluctuations.  If there are no quantum 
fluctuations, the spin configuration $\sigma_{i\rho K}$ does 
not depend on the Trotter number $K$ and $S_\rho (KL)=1$.
Quantum fluctuations reduce $S_\rho (KL)$  from unity.
Hence $1-S_\rho (KL)$ gives the scale of quantum fluctuations.

It is convenient to rewrite a function of $\tilde{\lambda}_{K\rho}$ as
\begin{eqnarray}
 & &G\{\tilde{\lambda}_{K\rho}\}
  =\int\prod_{(K\rho ,L\sigma )}
     {\rm d}q_{\rho\sigma}(KL)\prod_\rho \prod_{(KL)}{\rm d}S_{\rho}(KL)
     \nonumber\\
  & &\times \delta\left( q_{\rho\sigma}(KL)-\frac{1}{N}\sum_i
     \sigma_{i\rho K}\sigma_{i\sigma L} \right)
     \nonumber\\
  &  &\times\delta\left( S_{\rho}(KL)-\frac{1}{N}\sum_i
     \sigma_{i\rho K}\sigma_{i\rho L} \right) G\{\lambda_{K\rho}\}\ ,
     \label{Grho}
\end{eqnarray}
where $(KL)$ denotes an arbitrary combination of $K$ and $L$ including
$K=L$, and $(K\rho ,L\sigma )$ expresses an arbitrary
pair except for $\rho =\sigma$.
Using the Fourier representations of the delta functions,
(\ref{Grho}) is written as

\begin{full}
\begin{eqnarray}
  G\{\tilde{\lambda}_{K\rho}\}
  &=&\int\prod_{(K\rho ,L\sigma )}
     {\rm d}q_{\rho\sigma}(KL) {\rm d}r_{\rho\sigma}(KL)
     \prod_\rho \prod_{(KL)}{\rm d}S_{\rho}(KL){\rm d}t_{\rho}(KL)
     \nonumber\\
  & &\times\exp \left[-\frac{N\alpha\beta^2}{M^2}r_{\rho\sigma}(KL)
  q_{\rho\sigma}(KL)
      +\frac{\alpha\beta^2}{M^2}r_{\rho\sigma}(KL)
            \sum_i \sigma_{i\rho K}\sigma_{i\sigma L} \right.
       \nonumber\\
 & &\left. \hspace{1.1cm} -\frac{N\alpha\beta^2}{M^2}t_{\rho}(KL)S_{\rho}(KL)
      +\frac{\alpha\beta^2}{M^2}t_{\rho}(KL)
            \sum_i \sigma_{i\rho K}\sigma_{i\rho L} \right] 
            G\{\lambda_{K\rho}\}\ ,
     \label{Grho2}
\end{eqnarray}
\end{full}

\noindent
where we have ignored the overall constant.  

The average of the replicated $Z$ is thus 

\begin{full}
\begin{eqnarray}
  & &\ll Z^n \gg= \int \prod_{K\mu\rho} {\rm d}m_{K\mu\rho}
  \prod_{(K\rho ,L\sigma )}
      {\rm d}q_{\rho\sigma}(KL) {\rm d}r_{\rho\sigma}(KL)
      \prod_\rho \prod_{(KL)}{\rm d}S_{\rho}(KL){\rm d}t_{\rho}(KL)
                        \nonumber\\
   & &\times \exp \left[ -\frac{N\beta}{2M}
   \sum_{K\rho }\sum_{\mu\le s}m_{K\mu\rho}^2
      -\frac{N\alpha}{2}\sum_{K\rho } \log \lambda_{K\rho}\right.
      　　　　　　\nonumber\\
  & &\hspace{2cm} \left.-\frac{N\alpha\beta^2}{2M^2}\sum_{(K\rho ,L\sigma )}
                         r_{\rho\sigma}(KL)q_{\rho\sigma}(KL)
      -\frac{N\alpha\beta^2}{2M^2}\sum_\rho \sum_{(KL)}
                         t_{\rho}(KL)S_{\rho}(KL) \right]
                        \nonumber\\
     & &\times \ll \sum_\sigma \exp\left[ \frac{\beta}{M}\sum_{K\rho}
         \sum_{\mu \le s} m_{K\mu\rho}\sum_i \xi_i^\mu \sigma_{i\rho K}
      +B\sum_i \sum_{K\rho }\sigma_{i\rho K}\sigma_{i\rho, K+1}
        \right.
      \nonumber\\
      & &\hspace{2cm} +\left. \frac{\alpha\beta^2}{2M^2}\sum_i 
            \sum_{(K\rho ,L\sigma )}
                     r_{\rho\sigma}(KL)\sigma_{i\rho K} \sigma_{i\sigma L}
      +\frac{\alpha\beta^2}{2M^2}\sum_i \sum_\rho \sum_{(KL)}
        t_{\rho}(KL)  \sigma_{i\rho K} \sigma_{i\rho L} \right] \gg\ .
   \label{Zn2}
\end{eqnarray}
\end{full}

\noindent
If we write the above integrand as $\exp (-N\beta f)$, the saddle point
of the integral yields the free energy per spin $f$ as

\begin{full}
\begin{eqnarray}
   f&=&\frac{1}{2M}\sum_{K\rho }\sum_{\mu\le s}m_{K\mu\rho}^2
      +\frac{\alpha}{2\beta}\sum_{K\rho } \log \lambda_{K\rho}
            \nonumber\\
         & +&\frac{\alpha\beta}{2M^2}\sum_{(K\rho ,L\sigma )}
                         r_{\rho\sigma}(KL)q_{\rho\sigma}(KL)
      +\frac{\alpha\beta}{2M^2}\sum_\rho \sum_{(KL)}
                         t_{\rho}(KL)S_{\rho}(KL) 
                        \nonumber\\
  & -&T \ll\log \sum_\sigma \exp\left[ \frac{\beta}{M}\sum_{K\rho}
         \sum_{\mu \le s} m_{K\mu\rho} \xi^\mu \sigma_{\rho K}
      +B\sum_{K\rho }\sigma_{\rho K}\sigma_{\rho, K+1} \right.
      \nonumber\\
      & &\hspace{2cm}+\left.\frac{\alpha\beta^2}{2M^2} 
         \sum_{(K\rho ,L\sigma )}
                     r_{\rho\sigma}(KL)\sigma_{\rho K} \sigma_{\sigma L}
      +\frac{\alpha\beta^2}{2M^2}\sum_\rho \sum_{(KL)}
        t_{\rho}(KL)  \sigma_{\rho K} \sigma_{\rho L}\right] \gg\ .
  \label{ef1}
\end{eqnarray}
\end{full}

\noindent
We have used the self-averaging property of the free energy in dropping 
the site index $i$ in the above equation. The stationarity conditions 
of the free energy give the equations of state:
\begin{eqnarray}
  & &m_{K\mu\rho}=\ll\xi^\mu\langle \sigma_{\rho K}\rangle\gg\ ,
                  \label{mK}\\
  & &q_{\rho\sigma }(KL)=\ll 
                \langle\sigma_{\rho K}\rangle 
                \langle\sigma_{\sigma L}\rangle\gg\ ,
                   \label{qK}\\
  & &r_{\rho\sigma}(KL)=\frac{1}{\alpha}\sum_{\mu >s}
         \ll m_{K\mu\rho}m_{L\mu\sigma}\gg\ ,
                    \label{rK}\\
  & &S_{\rho}(KL)=\ll\langle\sigma_{\rho K}\sigma_{\rho L}\rangle\gg\ ,
                   \label{SK}\\
  & &t_{\rho}(KL)=\frac{1}{\alpha}\sum_{\mu >s}
                  \ll m_{K\mu\rho}m_{L\mu\rho}\gg\ .
                   \label{tK}
\end{eqnarray}
The physical meaning of the parameters $m_{K\mu\rho}$,
$q_{\rho\sigma }(KL)$ and $r_{\rho\sigma}(KL)$ is the
same as in the classical Hopfield model:
$m_{K\mu\rho}$ is the overlap, $q_{\rho\sigma }(KL)$ denotes 
the spin glass order parameter and $r_{\rho\sigma}(KL)$ 
represents the effects of uncondensed patterns.
The deviation of $S_{\rho}(KL)$ from unity reflects 
quantum fluctuations as mentioned before.  The last quantity
$t_{\rho}(KL)$ is for effects of uncondensed patterns in the same replica,
or in other words, the diagonal element of $r_{\rho\sigma}(KL)$.
%
\subsection{RS solution in the static approximation}
It is very difficult to solve the equations of state in their
general forms (\ref{mK})-(\ref{tK}).  We instead look for the solutions
in the replica symmetric (RS) subspace under the static
approximation.\cite{Usadel,static} We thus neglect the
dependence of the order parameters on the replica index
(the RS approximation) and the Trotter number (the static approximation):
\begin{eqnarray}
  &  &m_{K\mu\rho}= m_\mu\ ,
         \\
  &  &q_{\rho\sigma }(KL) = q\ ,
          \\
  &  &r_{\rho\sigma}(KL) = r\ ,
         \\
  &  &t_{\rho}(KL)= t\ ,
         \\
  &  &S_{\rho}(KL)=\left\{ 
        \begin{array}{ll}
             S& (K\ne L) \\
             1& (K=L)
         \end{array}\ .
      \right.
\end{eqnarray}
The stability of the replica symmetry will be considered later.
The static approximation is expected to give at least
qualitatively reliable results as long as the parameter
$\Delta$ representing quantum effects is not too large
as was the case of the SK model in a transverse field.\cite{Usadel,static}
The consistency of the RS and static approximations can be checked 
from the view point of another approximate method
as explained in $\S 3.3$ and Appendix A.

The free energy (\ref{ef1}) divided by $n$ (the number of replicas) is now
\begin{full}
\begin{eqnarray}
   f&=&\frac{1}{2}{\mib m}^2+\frac{\alpha}{2\beta n}\sum_{K\rho}
      \log \lambda_{K\rho}+\frac{\alpha\beta}{2M^2 }M^2 (n-1)rq
      +\frac{\alpha\beta}{2M^2 } M^2 tS
            \nonumber\\
    & &-\frac{T}{n} \ll\log\sum_\sigma \exp\left[\frac{\beta}{M}
      \sum_{K\rho}\sigma_{\rho K}\sum_{\mu\le s}m_\mu \xi^\mu
   +B\sum_{K\rho}\sigma_{\rho K}\sigma_{\rho ,K+1}  \right.
      \nonumber\\
       & &  \left. \hspace{1cm}+\frac{\alpha\beta^2}{2M^2}r\left\{
        \left(\sum_\rho \sum_K \sigma_{\rho K}\right)^2
        -\sum_\rho \left(\sum_K\sigma_{\rho K}\right)^2 \right\}
        +\frac{\alpha\beta^2}{2M^2}t\sum_\rho 
         \left(\sum_K\sigma_{\rho K}\right)^2
          \right]
        \gg\ .
   \label{ef2}
\end{eqnarray}
\end{full}

\noindent
The summation of $\log\lambda_{K\rho}$ appearing above is carried
out as follows. There are three values of the matrix elements of
$\{ \Lambda_{K\rho ,L\sigma}\} $ defined in (\ref{Lamb}),
\begin{eqnarray}
  & &-\frac{\beta }{M}q~~~(\rho\ne\sigma)\ ,
  \nonumber\\
  & &-\frac{\beta }{M}S~~~(\rho =\sigma, K\ne L)\ ,
  \nonumber\\
 & &1-\frac{\beta}{M}~~~(\rho =\sigma, K=L)\ ,
 \nonumber
\end{eqnarray}
 under the static approximation.
The eigenvalues of this matrix are easily found to be
\begin{eqnarray}
   \lambda_1&=&1-\frac{\beta}{M}-\frac{M-1}{M}\beta S
    -(n-1)\beta q\ ,
           \nonumber\\
   \lambda_2&=&1-\frac{\beta}{M}-\frac{M-1}{M}\beta S
    +\beta q\ ,
           \nonumber\\
    \lambda_3&=&1-\frac{\beta}{M}+\frac{\beta}{M}S\ ,
           \nonumber
\end{eqnarray}
with degeneracies $1, n-1$ and $(M-1)n$, respectively. Thus, we have

\begin{full}
\begin{eqnarray}
  \lim_{n\to 0}\frac{1}{n}\sum_{K\rho} \log\lambda_{K\rho}
  &=&\log \left(1-\frac{\beta}{M}-\frac{M-1}{M}\beta S+\beta q\right)
     \nonumber \\
    & &\hspace{3mm}
    -\frac{\beta q}{1-\frac{\beta}{M} -\frac{M-1}{M}\beta S+\beta q}
     +(M-1)\log \left(1-\frac{\beta}{M} +\frac{\beta}{M}S\right)
        \nonumber \\
  &\mathop{\longrightarrow}_{M\to \infty}&
    \log \left(1-\beta S+\beta q\right)
      -\frac{\beta q}{1-\beta S+\beta q}-\beta +\beta S\ .
      \label{log-l}
\end{eqnarray}
\end{full}

\noindent
The Gaussian integral enables us to decompose the
last term in the exponential appearing in (\ref{ef2}) into independent
replicas:

\begin{full}
\begin{eqnarray}
  &&\exp\frac{\alpha\beta^2}{2M^2} \left[ 
     r\left(\sum_\rho \sum_K \sigma_{\rho K}\right)^2
    +(t-r)\sum_\rho\left(\sum_K\sigma_{\rho K}\right)^2 \right]\nonumber\\
  &&=\int {\rm D}z \exp \left[
     \frac{\beta}{M}\sqrt{\alpha r}z\sum_{\rho K} \sigma_{\rho K}
      +\frac{\alpha\beta^2 (t-r)}{2M^2}
       \sum_\rho\left(\sum_K\sigma_{\rho K}\right)^2
        \right]\ ,
    \label{tr}
\end{eqnarray}
\end{full}

\noindent
where ${\rm D}z$ denotes the Gaussian measure 
${\rm e}^{-z^2/2}{\rm d}z/\sqrt{2\pi}$. Then, the summation 
over $\sigma_\rho$ in (\ref{ef2}) can be carried out 
independently for each $\rho$ as

\begin{full}
\begin{eqnarray}
  & &\log \sum_\sigma \exp \left[
   \frac{\beta}{M}\sum_K \sum_\rho \sigma_{\rho K}\sum_\mu m_\mu \xi^\mu
   +B\sum_K\sum_\rho \sigma_{\rho K}\sigma_{\rho ,K+1}\right.\nonumber\\
  & &\hspace{2.0cm}\left.
   +\frac{\alpha\beta^2}{2M^2}r\left(\sum_\rho \sum_K \sigma_{\rho K}\right)^2
      +\frac{\alpha\beta^2}{2M^2}(t-r)
       \sum_\rho\left(\sum_K\sigma_{\rho K}\right)^2
       \right] 
       \nonumber\\
   & &=\log \int{\rm D}z\left\{ \sum_\sigma \exp \left[
     \frac{\beta}{M}{\mib m}\cdot{\mib \xi}\sum_K \sigma_{\rho K}
      +B\sum_K \sigma_{\rho K}\sigma_{\rho ,K+1} \right. \right.
        \nonumber\\
        & &\hspace{4.0cm}\left.\left.
      +\frac{\beta}{M}\sqrt{\alpha r}z\sum_K \sigma_{\rho K}
      +\frac{\alpha\beta^2}{2M^2}(t-r) \left(\sum_K\sigma_{\rho K}\right)^2
       \right] \right\}^n 
       \nonumber\\
    & &\mathop{\approx}_{n\to 0} n\int{\rm D}z \log \sum_\sigma\exp \left[
     \frac{\beta}{M}{\mib m}\cdot{\mib \xi}\sum_K \sigma_{K}
      +B\sum_K \sigma_{K}\sigma_{K+1}\right.
        \nonumber\\
        & &\hspace{4.0cm}\left.
      +\frac{\beta}{M}\sqrt{\alpha r}z\sum_K \sigma_{K}
      +\frac{\alpha\beta^2}{2M^2}(t-r)\left(\sum_K\sigma_{K}\right)^2
       \right]  
        \nonumber\\
     & &=n \int{\rm D}z \log \sum_\sigma \int{\rm D}w\exp \left[
     \frac{\beta}{M}{\mib m}\cdot{\mib \xi}\sum_K \sigma_{K}
      +B\sum_K \sigma_{K}\sigma_{K+1}\right.
        \nonumber\\
        & &\hspace{4.8cm}\left.
      +\frac{\beta}{M}\sqrt{\alpha r}z\sum_K \sigma_{K}
      +\frac{\beta}{M}\sqrt{\alpha (t-r)}w\sum_K \sigma_{K} \right] 
       \nonumber\\
     & &\mathop{\longrightarrow}_{M\to \infty} n\int{\rm D}z\log\int{\rm D}w
      \, {\rm Tr} \exp \left[\beta \left\{ {\mib m}\cdot{\mib \xi}
                  +\sqrt{\alpha r}z
        +\sqrt{\alpha (t-r)}w\right\}\sigma_z +\beta\Delta \sigma_x \right] 
          \nonumber\\
     & &= n\int{\rm D}z\log\int{\rm D}w \, 2\cosh\beta
      \sqrt{\left[{\mib m}\cdot{\mib \xi}+\sqrt{\alpha r}z
      +\sqrt{\alpha (t-r)}w \right]^2
        +\Delta^2}\ .
      \label{tr2}
 \end{eqnarray}
\end{full}

\noindent
The total free energy is then obtained from 
(\ref{ef2}), (\ref{log-l}) and (\ref{tr2}) as

\begin{full}
\begin{eqnarray}
  f&=&\frac{1}{2}{\mib m}^2+\frac{\alpha}{2\beta} \left[
      \log(1-\beta S+\beta q)-\frac{\beta q}{1-\beta S+\beta q}-\beta (1-S)
      \right]
      +\frac{\alpha\beta}{2}(tS-rq)
            \nonumber\\
   & &-T\ll \int{\rm D}z\log\int{\rm D}w \, 2\cosh\beta
      \sqrt{\left[ {\mib m}\cdot{\mib \xi}+\sqrt{\alpha r}z
           +\sqrt{\alpha (t-r)}w\right]^2
        +\Delta^2} \gg\ .
   \label{ef3}
\end{eqnarray}
\end{full}
%
\subsection{Equations of state at $T=0$}
\label{sec.3.3}
Variation of the free energy (\ref{ef3}) gives the equations of
state for the order parameters.  Let us consider the case in which
only one of the patterns is retrieved, $m_\mu =\delta_{\mu 1}\cdot m$.
For brevity of expressions,  we introduce the following notations, 
%
\begin{eqnarray}
  g&=& m+\sqrt{\alpha r}z+\sqrt{\alpha (t-r)}w\ ,
    \nonumber \\
  u&=& \sqrt{g^2 +\Delta^2}\ ,
    \nonumber\\
   Y&=&\int {\rm D}w \cosh \beta u\ .
   \nonumber
\end{eqnarray}
%
The equations of state obtained by the variation of the free energy
(\ref{ef3}) with respect to $m, q, s, r$ and $t$ are, respectively,
\begin{eqnarray}
 & & m= \int {\rm D}z \, Y^{-1}\int {\rm D}w\, gu^{-1}\sinh \beta u\ ,
       \label{ma}\\
  &  &r= \frac{q}{(1-\beta S+\beta q)^2}\ ,
       \label{ra}\\
  &  & t= r+\frac{S-q}{1-\beta S+\beta q}\ ,
       \label{ta}\\
 &  &q= \int{\rm D}z \left(Y^{-1}\int {\rm D}w\, gu^{-1}\sinh \beta u
     \right)^2\ ,
       \label{qa}\\
  &  &S=\int{\rm D}z Y^{-1}\left(\int{\rm D}w\, g^2 u^{-2}\cosh \beta u\right.
    \nonumber\\
     & &   \hspace{1cm}
     \left. +T\Delta^2 \int{\rm D}w\, u^{-3}\sinh \beta u \right)\ .
       \label{sa}
\end{eqnarray}

Quantum effects play most important roles in the ground state
where thermal fluctuations are absent. We therefore only consider
the case $T=0$ hereafter.  The equations of state simplify
considerably in the $T=0$ limit because $q=S$ and
$t=r$ when $T=0$ as proved below.
For very small $T$, we find from (\ref{qa}) and (\ref{sa})
\begin{eqnarray}
 S&=&\int {\rm D}z \, Y^{-1}\int{\rm D}w\, g^2 u^{-2}\cosh \beta u
      \nonumber\\
    &\ge &\int {\rm D}z \, Y^{-1}\int{\rm D}w\, g^2 u^{-2}\sinh \beta u
      \nonumber\\
     &\ge &\int {\rm D}z \left( Y^{-1}\int{\rm D}w\, gu^{-1}\sinh \beta u
      \right)^2 =q\ .
      \nonumber
\end{eqnarray}
If we assume that $S$ is strictly larger than $q$ at $T=0$, 
(\ref{ta}) immediately leads to $t=r$ and also (\ref{ra}) to 
$r=0$ because $-\beta S+\beta q$ diverges. Then, $u$ is a constant
$(m^2+\Delta^2)^{1/2}$, and hence (\ref{qa}) and (\ref{sa}) 
imply $S=q$, a contradiction to the assumption $S>q$. 
Therefore, the equality $S=q$ holds in the ground state.

To derive the ground-state equation of state for $r$ from (\ref{ra}), 
we have to estimate the $T\to 0$ limit of $\beta(S-q)$. 
Comparison of (\ref{qa}) and (\ref{sa}) leads to
\begin{equation}
  \lim_{T\to 0}\beta (S-q)=
   \Delta^2 \int {\rm D}z\frac{1}
     {\left[\left(m+\sqrt{\alpha r}z\right)^2 +\Delta^2\right]^{3/2}}
     \equiv C\ .
\label{C}
\end{equation}
The equations of state at $T=0$ are thus written as
\begin{eqnarray}
  & &m=\int{\rm D}z
    \frac{m+\sqrt{\alpha r}z}
         {\sqrt{\left(m+\sqrt{\alpha r}z\right)^2 +\Delta^2}}
                 \label{eq-m}\ ,\\
  & &q=\int{\rm D}z\frac{\left(m+\sqrt{\alpha r}z\right)^2}
                  {\left(m+\sqrt{\alpha r}z\right)^2 +\Delta^2}
                  \label{eq-q}\ ,\\
   & &r=\frac{q}{(1-C)^2}\ .
                  \label{eq-r}
 \end{eqnarray}

The same equations of state can also be derived by a direct
mean-field analysis as explained in detail in Appendix A.
The reason why we have presented the replica analysis in 
the present section is three fold.  First, the replica
method gives the free energy by means of which we can
distinguish two different retrieval phases as explained
in the next subsection.  Second, the AT line, namely the stability
limit of the RS solution, can be determined by the replica
formalism as discussed in $\S$3.5. Lastly, it is useful
to confirm that two different methods lead to the 
same results so that we acquire confidence in the 
appropriateness of the present approximations.

\subsection{Phase diagram at $T=0$}
\begin{figure}
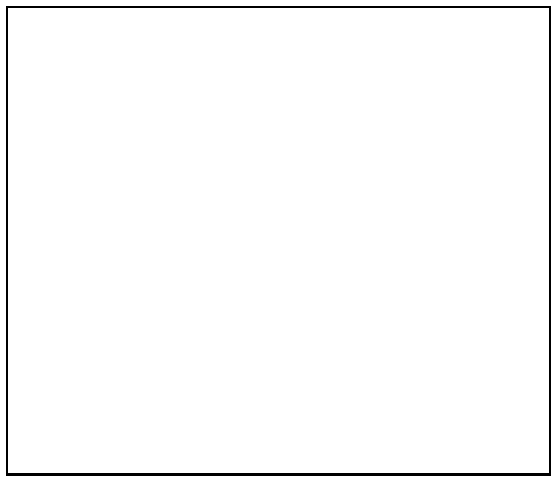

  \figureheight{6cm}
  \caption{
  The ground-state phase diagram of the Hopfield model in a 
transverse field. ``R-I" stands for the retrieval phase in which 
the retrieval states are the global minima, and ``R-II" denotes the 
retrieval phase where the retrieval states are the local minima. 
The dashed line represents the AT line. 
  }
  \label{Phase-diagram}
\end{figure}

To draw the phase diagram, the equations of state (\ref{eq-m}),
(\ref{eq-q}) and (\ref{eq-r}) have been solved
numerically.  The result is drawn in Fig.~\ref{Phase-diagram}.
The asymptotic form of the overlap order parameter around the
The three phases are characterized by the relations
$m=q=0$ (paramagnetic, P), $m=0, q>0$ (spin glass, SG)
and $m\ne 0, q>0$ (retrieval, R-I and R-II), respectively.
The retrieval phase is separated into two parts,
one with $f_{\rm R}<f_{\rm SG}$ (R-I) and the other with
$f_{\rm SG}<f_{\rm R}$ (R-II), where $f_{\rm R}$ is the
free energy of the retrieval state and $f_{\rm SG}$
is that of the spin glass state.

The transition between the paramagnetic and the
spin glass phases is of second order, and so
the shape of the boundary can be determined
analytically by expansion of the equation of state
(\ref{eq-q}).  The result is
\begin{equation}
   \Delta =1+\sqrt{\alpha}\ .
     \label{SGp}
\end{equation}
This is exactly the same relation as the corresponding classical 
phase boundary if we replace $\Delta$ by $T$.\cite{AGS2}
The other phase transitions (between SG and R-II and
between R-I and R-II) are both of first order. Therefore, 
it is in general impossible to obtain the analytic 
expressions of these phase boundaries. However,
when $\alpha$ is very small, we can determine
the asymptotic forms of the phase boundaries by expansions of
the equations of state and the free energy as in the case of
the classical model.\cite{AGS2}  Details are described in 
Appendix \ref{asymp}.
The result is 
\begin{equation}
  \Delta\simeq1-1.95\sqrt{\alpha}
  \label{fnbou1}
\end{equation}
for the transition between the SG and R-II phases, and 
\begin{equation}
  \Delta\simeq 1-\frac{33}{16}\sqrt{\alpha}=1-2.0625\sqrt{\alpha}
  \label{fnbou2}
\end{equation}
between the R-I and R-II regions. The relation (\ref{fnbou1}) 
exactly agrees with that of the classical case if we replace 
$\Delta$ by $T$.\cite{AGS2} The relation (\ref{fnbou2}) 
shows that the R-I region of this model in the vicinity of 
($\alpha=0$, $\Delta=1$) is wider than that of the classical 
case, $T\simeq 1-2.6\sqrt{\alpha}$,\cite{AGS2} though the 
deviation from unity is also proportional to $\sqrt{\alpha}$. 

When $\Delta=0$, the R-II phase changes into the SG 
phase at $\alpha =0.1379$ as it should.\cite{Steffan}
The boundary between the SG and R-II phases is
slightly reentrant in the low-temperature region. That is,
for a fixed $\alpha$ slightly larger than 0.1379,
the SG phase once changes to the R-II phase as $\Delta$ is
decreased but the SG phase once again becomes stable
for very small $\Delta$.  The same is true for the
boundary between the R-I and R-II regions.  The reentrance
in these two cases is observed also in the classical model.
It should be noted, however, that the effect of replica 
symmetry breaking treated in the next subsection obscures
the significance of reentrance within the RS solution
near $\alpha =0.1379$, again the same situation 
as in the classical model.\cite{Steffan}
%
\subsection{AT line}
The stability of the replica symmetric solution 
against replica symmetry breaking can be checked 
following the standard procedure.\cite{AGS2,AT}
We keep the static approximation intact and investigate
only the effects of replica symmetry breaking. 
Calculations are somewhat involved but straightforward. 
Details are given in Appendix \ref{AT}.
The stability limit of the replica symmetric 
solution (the AT line) is found to be given by 
\begin{equation}
  q=\alpha r\Delta^4\int{\rm D}z
    \left[\left(m+\sqrt{\alpha r}z\right)^2+\Delta^2\right]^{-3}\ .
  \label{AT-line}
\end{equation}
This AT line is also drawn in the phase diagram in Fig.~\ref{Phase-diagram}.

\begin{figure}
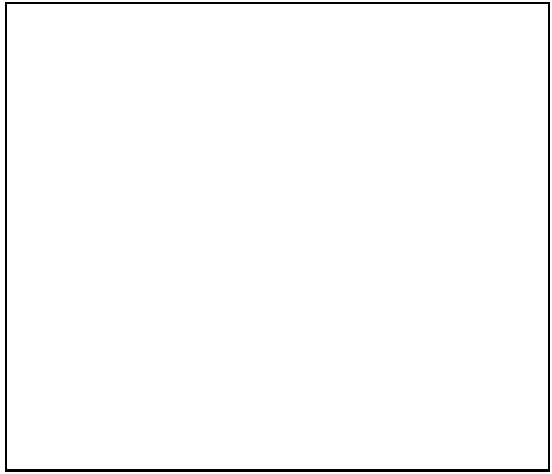

  \figureheight{6cm}
  \caption{The phase diagram near $\alpha=0.1379$ and  $\Delta =0$.
  }
  \label{Enlarged}
\end{figure}

The region near $\alpha=0.1379$ and  $\Delta =0$ is drawn enlarged
in Fig.~\ref{Enlarged}.
The AT line merges with the boundary between the R-I and 
SG phases at $\alpha\simeq 0.1378$ and $\Delta\simeq 0.022$.
Since the replica-symmetric solution is unstable below this line, 
the reentrant behavior  mentioned in the previous subsection 
may be an artifact of the replica symmetric solution.

\section{Summary and Discussion}
The Hopfield model in a transverse field has been introduced 
and solved. For a finite number of embedded patterns,
we have used the method of the Trotter decomposition
to reduce the quantum problem to a classical form. 
We found that the solutions
of the equations of state in the ground state
have quite similar properties to those of the 
classical model at finite temperatures.    

If the number of embedded patterns is
proportional to the system size, it is necessary to employ the
replica technique, in addition to the Trotter decomposition, to
trace out the quenched randomness. We cannot obtain the full 
exact solution and have to appeal to the replica symmetric (RS) and
static approximations.  The stability analysis of the RS solution leads
to the AT line, above which the RS solution is at least locally stable.
On the other hand, the static approximation probably does not give the
exact solutions for all  values of the parameters except for the 
limits of $\Delta\to 0$ (the classical model) and $\alpha \to 0$ 
(the finite-$p$ model).
Nevertheless, the experience in the transverse SK model\cite{Usadel}
suggests that the static approximation is expected to capture the 
qualitative features of spin systems in transverse fields. 
The resulting phase diagram has three phases, the retrieval, 
spin glass and the paramagnetic phases.  The shapes of phase 
boundaries turn out to be almost the same as those of the 
classical model if we replace $\Delta$ by $T$.

We have found that the quantum fluctuations have almost the 
same effects as thermal fluctuations in the Hopfield model.
The uncertainties in signal transmission at a synapse have 
conventionally been treated in model analyses in terms of 
thermal fluctuations. The results of the present study show that 
we may instead consider quantum fluctuations without changing 
conclusions on the macroscopic behavior of the network.

The above-mentioned fact means that quantum uncertainties do not lead to 
many possible alternative values of macroscopic variables but rather 
cause only quantitative deteriorations as thermal fluctuations do.
Quantum uncertainties at microscopic levels are far from the 
source of the simultaneous existence of macroscopically
distinguishable states.  This point seems to be disregarded
in some of the arguments concerning the significance of 
quantum mechanics in the functioning of the brain.\cite{Beck_Eccles}

\acknowledgement
One of the present authors (Y.\ N.) is grateful for the financial support
of the Japan Society for the Promotion of Science for Japanese Junior 
Scientists. Numerical calculations were performed on FACOM VPP 500 
at the Institute for Solid State Physics, University of Tokyo. 

\appendix
\section{Equation of State by the Mean-Field Theory}
\label{Gestzi}
In this Appendix we derive the equations of state
(\ref{eq-m}), (\ref{eq-q})  and (\ref{eq-r}) by a
direct mean-field analysis.\cite{Gestzi}
The Hamiltonian of the Hopfield model
(\ref{Hamiltonian}) with the Hebb rule (\ref{Hebb})
is approximated by the following single-site
effective Hamiltonian:
\begin{equation}
  {\cal H}_i = -\sigma_i^z\sum_\mu \xi_i^\mu m_\mu -\Delta \sigma_i^x\ ,
  \label{Hi}
\end{equation}
%
where $m_\mu$ stands for the overlap
\begin{equation}
     m_\mu = \frac{1}{N}\sum_j \xi_j^\mu \langle \sigma_j^z \rangle\ .
 \label{mmu}
\end{equation}
The thermal expectation value of $\sigma_j^z$ appearing in (\ref{mmu})
can be calculated easily under the Hamiltonian (\ref{Hi}) to give
\begin{equation}
   \langle \sigma_i^z \rangle = \frac{u_i}{\sqrt{u_i^2+\Delta^2}}
      \tanh \beta\sqrt{u_i^2 +\Delta^2}\ ,
 \label{sigmai}
\end{equation}
where
\begin{equation}
   u_i = \sum_\mu m_\mu \xi_i^\mu\ .
 \label{ui}
\end{equation}
Let us now assume that a single pattern, the first one ($\mu =1$) 
for instance, is retrieved. Then, $m_1 \equiv m$ is of order unity 
and all other $m_\mu $'s are of order $1/\sqrt{N}$. The equation 
of state (\ref{mmu}) for $\mu =1$ is then written as
\begin{eqnarray}
 m&=&\frac{1}{N}\sum_i \frac{m+\sqrt{\alpha r}z_i}
      {\sqrt{\left(m+\sqrt{\alpha r}z_i \right)^2 +\Delta^2}}
      \nonumber\\
  & &\times\tanh\beta\sqrt{\left(m+\sqrt{\alpha r}z_i \right)^2 +\Delta^2}\ ,
  \label{man}
\end{eqnarray}
where
\begin{equation}
  \sqrt{\alpha r}\, z_i = \sum_{\mu\ge 2}
       m_\mu \xi_i^\mu \xi_i^1\ .
 \label{zi}
\end{equation}
Since $z_i$ is proportional to the sum of extensively 
many random variables, it is not unreasonable to
assume that $z_i$ is a Gaussian random variable
according to the central limit theorem.
Strictly speaking, the terms appearing in the definition
(\ref{zi}) are mutually correlated through the 
$\xi$-dependence of $m_\mu$ and the central limit theorem
is not applicable.  Nevertheless, the
results obtained under this approximation of
Gaussian distribution agree with those from the replica
method in the case of the classical Hopfield model.\cite{Gestzi}
We thus adopt the same approach in the present quantum case.

The average value of $z_i$ is expected to be vanishing because
$\xi_i^\mu\xi_i^1$ is $\pm 1$ with equal probability.
The variance of $z_i$ is unity if we define $r$ in the following way:
\begin{equation}
    \ll \left( \sum_{\mu\ge 2}m_\mu \xi_i^\mu \xi_i^1\right) ^2 \gg
    = \ll\sum_{\mu\ge 2}m_\mu^2 \gg =\alpha r\ .
 \label{ar}
\end{equation}
Then, in the thermodynamic limit, the summation 
over $i$ in (\ref{man}) reduces to the average 
over the Gaussian distribution as
\begin{eqnarray}
  m &=&\int {\rm D}z \frac{m+\sqrt{\alpha r}z}
      {\sqrt{\left(m+\sqrt{\alpha r}z \right)^2 +\Delta^2}}
      \nonumber\\
       &\times &
        \tanh\beta\sqrt{\left(m+\sqrt{\alpha r}z \right)^2 +\Delta^2}\ ,
\label{Gm}
\end{eqnarray}
where
${\rm D}z$ denotes the Gaussian measure 
${\rm e}^{-z^2/2}{\rm d}z/\sqrt{2\pi }$.

To derive the equation of state for $r$, we need to evaluate $m_\nu$ 
for $\nu \ge 2$ according to (\ref{ar}).  For this purpose 
we decompose $u_i$ defined in (\ref{ui}) into
three parts by extracting contributions from $\mu =1$ and $\mu = \nu$:
\begin{equation}
  u_i \xi_i^1 = m+\sqrt{\alpha r}z_i+\eta_i^\nu m_\nu\ ,
  \label{ui2}
\end{equation}
%
with $\eta_i^\nu =\xi_i^\nu \xi_i^1$. Although the definition 
of $z_i$ here does not include the contribution 
from $\mu =\nu$ in contrast to (\ref{zi}), this 
difference does not affect the results because the only property
of $z_i$ we use is its Gaussian distribution, which remains the
same if we drop a single term from (\ref{zi}).  It should be noted 
here that $m_\nu$ is of order $1/\sqrt{N}$ and is quite small 
compared to the other two terms in (\ref{ui2}).

We insert (\ref{ui2}) into the right hand side of
(\ref{mmu}) with (\ref{sigmai}) taken into account
and expand the result to first order of $m_\nu$ to find

\begin{full}
\begin{eqnarray}
  m_\nu &=&\frac{1}{N}\sum_i \frac{v_i}{\sqrt{v_i^2 +\Delta^2}}
                      \tanh\beta\sqrt{v_i^2 +\Delta^2}  
           \nonumber\\
       &+&\frac{m_\nu}{N}\sum_i 
            \left[ \frac{\Delta^2}{(v_i^2 +\Delta^2)^{3/2}} 
                      \tanh\beta\sqrt{v_i^2 +\Delta^2}  
       +\frac{\beta v_i^2}{v_i^2 +\Delta^2} 
                     {\rm sech}^2\beta\sqrt{v_i^2 +\Delta^2}  \right]\ ,
            \label{mnu}
\end{eqnarray}
\end{full}

\noindent
with $v_i =\eta_i^\nu \left(m+\sqrt{\alpha r}z_i \right)$. 
We now define $q$ and $S$ as
\begin{eqnarray}
  q&=& \frac{1}{N}\sum_i\frac{v_i^2}{v_i^2 +\Delta^2}
                      \tanh^2 \beta\sqrt{v_i^2 +\Delta^2}  
             \nonumber \\
       &=& \int {\rm D}z \frac{v^2}{v^2 +\Delta^2}
                       \tanh^2 \beta\sqrt{v^2 +\Delta^2}  
             \label{Gq} \\
  S&=& 1-\int {\rm D}z \left[ \frac{\Delta^2}{v^2 +\Delta^2} \right.
       \nonumber\\
   & &-\left.\frac{T\Delta^2}{(v^2 +\Delta^2)^{3/2}} 
                          \tanh \beta\sqrt{v^2 +\Delta^2}\right]\ ,
              \label{GS}
\end{eqnarray}
with $v=m+\sqrt{\alpha r} z$.  Equation (\ref{mnu}) 
can then be solved for $m_\nu$ as
\begin{eqnarray}
   m_\nu &=& \frac{1}{1-\beta S+\beta q} 
   \nonumber\\
   &\times &
   \frac{1}{N}
      \sum_i \frac{v_i}{v_i^2+\Delta^2}
               \tanh \beta\sqrt{v_i^2 +\Delta^2}\ .
     \label{mnu2}
\end{eqnarray}
{}From (\ref{mnu}) and (\ref{Gq}), we find
\begin{eqnarray}
    & & (1-\beta S+\beta q)^2 \ll\sum_{\nu\ge 2} m_\nu^2 \gg
        \nonumber \\
    &=& \frac{p-1}{N^2}\sum_i \frac{v_i^2}{v_i^2 +\Delta^2}
           \tanh^2 \beta\sqrt{v_i^2 +\Delta^2}
           \nonumber\\
     &=& \alpha q\ .   \nonumber
\end{eqnarray}
Equation (\ref{ar}) and the above relation lead to
\begin{equation}
   r=\frac{q}{(1-\beta S +\beta q)^2}\ .
  \label{Gr}
\end{equation}
It is straightforward to check that the zero-temperature limits 
of (\ref{Gm}), (\ref{Gq}), (\ref{GS}) and (\ref{Gr}) agree
with (\ref{C}), (\ref{eq-m}), (\ref{eq-q}) and (\ref{eq-r})
derived by the replica method, respectively. 

\section{Phase Boundaries Near $\alpha =0$}
\label{asymp}
In this Appendix we derive the asymptotic forms of phase boundaries
around the point ($\alpha =0, \Delta =1$).  The first boundary is 
between the SG and R-II phases and the second is between the
R-I and R-II phases. 

To derive the boundary between the SG and R-II phases, we expand
the right-hand side of the equation of state (\ref{eq-m}) assuming
$m$ and $r$ are small:
\[
  m=\frac{1}{\Delta }\left[ m-\frac{1}{2\Delta^2}(m^3+3m\alpha r)
  \right]\ .
  \]
Let us write $1-\Delta =\epsilon$.  Then, the above 
equation is approximated to order $\epsilon$ as
\begin{equation}
  m^2 +3\alpha r =2\epsilon\ .
    \label{eq-ma}
\end{equation}
Similarly, the equations of state for $C$, $q$ and $r$ 
((\ref{C}), (\ref{eq-q}) and (\ref{eq-r})) behave asymptotically as
\begin{eqnarray}
  \label{eq-Ca}
  C&=& (1+\epsilon )\left[ 1-\frac{3}{2}(m^2 +\alpha r)\right]\ ,\\
  \label{eq-qa}
  q&=& m^2+\alpha r\ ,\\
  r&=& \frac{4(m^2 +\alpha r)}{(2\epsilon -3m^2 -3\alpha r)^2}\ .
     \label{eq-ra}
\end{eqnarray}
To erase $r$ from (\ref{eq-ma}) and (\ref{eq-ra}), 
we define $x=m^2+\alpha r$ and write the above equations as
\begin{eqnarray}
  x&=& m^2 +\frac{4\alpha x}{(3x-2\epsilon )^2}\ ,
     \label{x1}\\
  x&=& \frac{2}{3}(\epsilon +m^2)\ .
     \label{x2}
\end{eqnarray}
{}From these equations we obtain
\begin{equation}
  \frac{2}{3}(\epsilon +m^2)=m^2+\frac{2\alpha (\epsilon +m^2)}{3m^4}\ .
  \label{coneq}
\end{equation}
Using new parameters $y$ and $\tau$ defined by 
\begin{eqnarray}
  \label{y}
  &&y=m^2/\sqrt{\alpha}\ ,\\
  &&\tau =\epsilon /\sqrt{\alpha}\ ,
  \label{tau}
\end{eqnarray}
(\ref{coneq}) is expressed as 
\begin{equation}
  \frac{1}{2}y^3 -\tau y^2 +y+\tau =0\ ,
  \label{boueq1}
\end{equation}
which coincides with (5.12) of Ref.\ \citen{AGS2}, though the 
definition of $y$ is not the same. Since bifurcation occurs 
on the SG--R-II phase boundary, the derivative of this 
equation is also vanishing on this boundary, that is, 
\begin{equation}
  \frac{3}{2}y^2-2\tau y+1=0\ .
  \label{dereq}
\end{equation}
Eliminating $\tau$ from (\ref{boueq1}) and (\ref{dereq}), we have 
\[
  \frac{\frac{1}{2}y^3+y}{y^2-1}=\frac{\frac{3}{2}y^2+1}{2y}\ , 
\]
and therefore 
\[
  y^2=\frac{5+\sqrt{33}}{2}\mbox{\ \ \ and\ \ \ }\tau=1.95\cdots\ ,
\]
or equivalently, 
%
\[
  \Delta\simeq 1-1.95\sqrt{\alpha}\ .
\]

The boundary between the R-I and R-II phases is the line on which 
the free energy of the solution with $m\neq 0$ and that of the 
solution with $m=0$ are equal to each other. We first derive the 
explicit form of the free energy at $T=0$. All we have to do is 
to take the limit $\beta\to\infty$ in the expression of the free 
energy at finite temperatures (\ref{ef3}). However, we should be 
very careful in taking this limit because both $(S-q)$ and $(t-r)$ 
are of order $T$, as shown in (\ref{ta}) and (\ref{C}). In order to 
evaluate the last term of (\ref{ef3}), $-T\ll\cdots\gg$, we expand this 
term up to first order of $(t-r)$ for the single retrieval case to obtain 

\begin{full}
\begin{eqnarray}
  && -T\ll\cdots\gg\nonumber\\
  &&=-T\int{\rm D}z\log\int{\rm D}w\ 2\cosh\left[\beta
     \sqrt{\left(m+\sqrt{\alpha r}z\right)^2+\Delta^2}
     \left(1+\frac{\left(m+\sqrt{\alpha r}z\right)\sqrt{\alpha(t-r)}w}
                  {\left(m+\sqrt{\alpha r}z\right)^2+\Delta^2}
           +\cdots\right)\right]\nonumber\\
  &&=-T\left\{
     \int{\rm D}z\log\left[
          \exp\left(\beta\sqrt{\left(m+\sqrt{\alpha r}z
                                     \right)^2+\Delta^2}\right)
          \int{\rm D}w\exp\left(\beta
          \frac{\left(m+\sqrt{\alpha r}z\right)\sqrt{\alpha(t-r)}w}
               {\sqrt{\left(m+\sqrt{\alpha r}z\right)^2+\Delta^2}}
                                \right)\right.\right.\nonumber\\
  && \hspace{2.8cm}\left.\left.
         +\exp\left(-\beta\sqrt{\left(m+\sqrt{\alpha r}z
                                      \right)^2+\Delta^2}\right)
          \int{\rm D}w\exp\left(-\beta
          \frac{\left(m+\sqrt{\alpha r}z\right)\sqrt{\alpha(t-r)}w}
               {\sqrt{\left(m+\sqrt{\alpha r}z\right)^2+\Delta^2}}
                                \right)
                           \right]\right\}\nonumber\\
  &&=-T\left[
     \int{\rm D}z\log 2\cosh\left(
         \beta\sqrt{\left(m+\sqrt{\alpha r}z\right)^2+\Delta^2}\right)
     +\frac{1}{2}\beta^2\alpha(t-r)
      \int{\rm D}z\frac{\left(m+\sqrt{\alpha r}z\right)^2}
                       {\left(m+\sqrt{\alpha r}z\right)^2+\Delta^2}
             \right]\nonumber\\
  &&\mathop{\longrightarrow}_{T\to 0}
     -\int{\rm D}z\sqrt{\left(m+\sqrt{\alpha r}z\right)^2+\Delta^2}
     -\frac{1}{2}\alpha\beta(t-r)q\ .
\end{eqnarray}
\end{full}

\noindent
Substituting this formula to (\ref{ef3}) 
and taking the limit $T\to 0$, we have 
\begin{eqnarray}
  f&=&\frac{1}{2}m^2-\frac{\alpha}{2}\left(1+\frac{C}{1-C}q-rC\right)
       \nonumber\\
   & & -\int{\rm D}z\sqrt{\left(m+\sqrt{\alpha r}z\right)^2+\Delta^2}\ .
  \label{fgs}
\end{eqnarray}

Next, we expand this expression with respect to $m^2$ and 
$\epsilon=1-\Delta$. Note that the last term of (\ref{fgs}) is 
not multiplied by a smallness parameter $\alpha/2$ in contrast to 
the preceding term, and thus we have to calculate up to second 
order of $\epsilon$ in this term. We aim to express the free energy 
only using the parameters $y$ and $\tau$ as (\ref{boueq1}). 
For this purpose, we write all the parameters in terms of 
$m^2$ and $\epsilon$. In the solution with $m\neq 0$, 
(\ref{eq-ma}), (\ref{eq-qa}) and (\ref{eq-Ca}) reduce to 
\begin{eqnarray}
  \label{exp1}
  &&q=\frac{2}{3}(m^2+\epsilon)\ ,\\
  &&\alpha r=\frac{2}{3}\epsilon-\frac{1}{3}m^2\ ,\\
  &&C=1+\epsilon-(m^2+\epsilon)\ .
\end{eqnarray}
On the other hand, in the solution with $m=0$, we use 
(\ref{eq-r}) instead of (\ref{eq-ma}). Substituting 
the relation $q=\alpha r$ to (\ref{eq-r}), we have 
$(1-C)^2=\alpha$. Because the term 
\[
  \frac{\alpha}{2\beta}\log(1-\beta S+\beta q)
  \mathop{\longrightarrow}_{T\to 0}\frac{\alpha}{2\beta}\log(1-C)
\]
exists in the original expression of the free energy (\ref{ef3}), 
the condition $C<1$ should be satisfied, and we obtain 
\begin{eqnarray}
  &&q=\alpha r=\frac{2}{3}\left(\epsilon+\sqrt{\alpha}\right)\ ,\\
  &&C=1-\sqrt{\alpha}\ .
  \label{exp5}
\end{eqnarray}
Using (\ref{y}), (\ref{tau}) and (\ref{fgs})--(\ref{exp5}), 
after long but straightforward calculations we have 
\begin{equation}
  f(m)-f(0)=\frac{\alpha}{6}\left[
            \frac{1}{2}y^2-2\tau\left(y+\frac{1}{y}\right)+4\tau+1\right]=0\ .
\end{equation}
Solving this equation together with (\ref{boueq1}), we obtain 
\[
  y=3\mbox{\ \ \ and\ \ \ }\tau=\frac{33}{16}=2.0625\ ,
\]
or equivalently, 
\[
  \Delta\simeq 1-2.0625\sqrt{\alpha}\ .
\]

%
\section{AT Line}
\label{AT}
We derive the expression of the AT line (\ref{AT-line}) at $T=0$.
We follow the Appendix B of Ref.\citen{AGS2} in the first half
of the argument.
To obtain the second derivatives of the free energy
by the variables $q_{\rho\sigma}$ and $r_{\rho\sigma}$
under the static approximation, we retain only the relevant part
in (\ref{ef1}):

\begin{full}
\begin{eqnarray}
   f&=&\frac{\alpha}{2\beta} \sum_{K\rho} \log \lambda_{K\rho}
    +\frac{\alpha\beta}{2}\sum_{\rho\sigma} r_{\rho\sigma}q_{\rho\sigma}
    \nonumber\\
    & &-T\ll \log \sum_\sigma \exp \left[\frac{\beta}{M}\sum_{\rho\mu}
        m_{\mu\rho}\xi^\mu \sum_K \sigma_{\rho K}\right.
      \nonumber\\
    & & \left. \hspace{4cm}+B\sum_{K\rho}\sigma_{\rho K}\sigma_{\rho ,K+1}
      +\frac{\alpha\beta^2}{2M^2}\sum_{\rho\sigma}r_{\rho\sigma}
        \sum_{KL}\sigma_{\rho K}\sigma_{\sigma L} \right] \gg\ .
         \label{ef-A1}
\end{eqnarray}
\end{full}

\noindent
We have set $r_{\rho\rho}=t_\rho$ in anticipation of
the ground state relation $r=t$ at the replica-symmetric point as explained
in $\S$3.3.
    
The Hessian, or the second-derivative matrix of the free energy, is
written as
\begin{equation}
  \left[
   \begin{array}{ll}
     A^{\alpha\beta, \gamma\delta} & \delta^{\alpha\beta, \gamma\delta} \\
     \delta^{\alpha\beta, \gamma\delta} &B^{\alpha\beta, \gamma\delta} 
    \end{array}
  \right]\ ,
         \label{Hessian}
\end{equation}
where 
\begin{eqnarray}
  & &A^{\alpha\beta, \gamma\delta} =
     \frac{\partial^2 f}{\partial q_{\alpha\beta}\partial q_{\gamma\delta}}\ ,
         \label{A-A}\\
  & &B^{\alpha\beta, \gamma\delta}=
     \frac{\partial^2 f}{\partial r_{\alpha\beta}\partial r_{\gamma\delta}}\ ,
         \label{A-B}\\
   & &\delta^{\alpha\beta, \gamma\delta}=
     \frac{\partial^2 f}{\partial q_{\alpha\beta}\partial r_{\gamma\delta}}\ .
         \label{A-delta}
\end{eqnarray}
There is no difficulty in evaluating the derivatives 
written above by taking into
account that $\{\lambda_{K\rho}\}$ is a set of eigenvalues of the 
$Mn\times Mn$ matrix (\ref{Lamb}) under the static approximation.
Note that the replica-index dependence of the matrix elements
should be kept untouched until differentiations are carried out.
The result is almost the same as that of the classical 
Hopfield model.\cite{AGS2} For the submatrix $A$,
\begin{eqnarray}
  & &A^{\rho\sigma ,\rho\sigma} =-\alpha\beta
     \left(C_{\rho\rho}^2+C_{\rho\sigma}^2\right)\ ,
     \\
  & &A^{\rho\sigma ,\rho\gamma} =-\alpha\beta
     \left(C_{\rho\rho}C_{\rho\sigma}+C_{\rho\sigma}^2\right)\ ,
     \\
  & &A^{\rho\sigma ,\gamma\delta} =-\alpha\beta
     \left(2C_{\rho\sigma}^2\right)\ ,
\end{eqnarray}
where
\begin{eqnarray}
  C_{\rho\sigma}&=&\frac{\beta q}{\left[1-\beta (S-q)\right]^2}
    \label{Crs}\ ,\\
  C_{\rho\rho}&=&C_{\rho\sigma}+\frac{1}{1-\beta (S-q)}\ .
    \label{Crr}
\end{eqnarray}
The elements of the submatrix $B$ are expressed as
\begin{eqnarray}
  &&B^{\rho\sigma ,\rho\sigma} =-\alpha^2 \beta^3
    \ll \langle \sigma_{\rho K}\sigma_{\rho L} \rangle^2
         -\langle \sigma_{\rho K} \rangle^4 \gg\ ,\\
   &&B^{\rho\sigma ,\rho\gamma} 
   \nonumber\\
   &&=-\alpha^2 \beta^3
    \ll \langle \sigma_{\rho K}\sigma_{\rho L} \rangle
         \langle \sigma_{\rho K}\rangle^2
         -\langle \sigma_{\rho K} \rangle^4 \gg\ ,\\
  & & B^{\rho\sigma ,\gamma\delta} =-\alpha^2 \beta^3
    \ll \langle \sigma_{\rho K}\rangle^4
         -\langle \sigma_{\rho K} \rangle^4 \gg =0\ ,
\end{eqnarray}
where $\langle\cdots\rangle$ denotes the average with respect to 
the weight appearing as the exponential function in (\ref{ef-A1}). 
This average should be evaluated at the replica-symmetric point 
to check the stability of the replica-symmetric solution against
replica-symmetry breaking.  The off-diagonal element of the 
block matrix (\ref{Hessian}) is given as
\begin{equation}
   \delta^{\sigma\rho ,\gamma\delta}=
     \alpha\beta \left(\delta_{\rho\gamma}\delta_{\sigma\delta}
       +\delta_{\rho\delta}\delta_{\sigma\gamma} \right)\ .
     \label{A-D}
\end{equation}

We follow the usual procedure to find 
the eigenvalue of the replicon mode, 
\begin{eqnarray}
  q_{\rho\sigma}&=&q+\eta_{\rho\sigma}\ ,
    \label{A-Repq}\\
  r_{\rho\sigma}&=&r+x\eta_{\rho\sigma}\ ,
    \label{A-Repr}
\end{eqnarray}
where
\begin{eqnarray}
    \eta_{\rho\sigma}&=& \eta ~~~(\rho ,\sigma\ne 1,2)\ ,
             \nonumber\\
    \eta_{1\rho}&=&\eta_{2\rho}=\frac{1}{2}(3-n)\eta ~~~(\rho\ne 1,2)\ ,
              \nonumber\\
    \eta_{12}&=&\frac{1}{2}(2-n)(3-n)\eta\ ,
              \nonumber\\
    \eta_{\rho\rho} &=&0\ .
              \nonumber
\end{eqnarray}
The eigenvalue equations are then given by 
\begin{eqnarray}
  & &\sum_{\gamma\delta}\left(A^{\rho\sigma ,\gamma\delta}
        +x\delta^{\rho\sigma ,\gamma\delta }\right)\eta_{\gamma\delta}
          =\lambda\eta\ ,
            \label{A-eigen1}\\
   & &\sum_{\gamma\delta}\left(xB^{\rho\sigma ,\gamma\delta}
        +\delta^{\rho\sigma ,\gamma\delta }\right)\eta_{\gamma\delta}
          =\lambda x\eta
            \label{A-eigen2}\ ,
\end{eqnarray}
with $\rho ,\sigma \ne 1,2$.  Substitution of the expressions of
$A^{\rho\sigma ,\gamma\delta}$, $B^{\rho\sigma ,\gamma\delta}$,
 $\delta^{\rho\sigma ,\gamma\delta}$ and $\eta_{\gamma\delta}$ 
 into these equations leads
 to
\begin{eqnarray}
   & &x=\tilde{\lambda}+\frac{1}{\left[1-\beta (S-q)\right]^2}\ ,
       \nonumber\\
   & &x\left[\alpha\beta^2 \ll 
       \left(\langle \sigma_{\rho K}\sigma_{\rho L}\rangle
      -\langle \sigma_{\rho K} \rangle^2 \right)^2 \gg +\tilde{\lambda}
       \right]=1\ ,
       \nonumber
\end{eqnarray}
with $\tilde{\lambda}=\lambda /\alpha\beta$. 
This set of equations is solved for the eigenvalues as
\[
  \tilde{\lambda}_{\pm}=-\frac{1}{2}(u+v)\pm
          \sqrt{\frac{1}{4}(u+v)^2+1-uv}\ ,
\]
where
\begin{eqnarray}               
     u&=&\alpha\beta^2 \ll \left(\langle \sigma_{\rho K}\sigma_{\rho L}\rangle
     -\langle \sigma_{\rho K} \rangle^2 \right)^2 \gg\ ,
               \nonumber\\
     v&=&\frac{1}{\left[1-\beta (S-q)\right]^2}=\frac{r}{q}\ .
               \nonumber
\end{eqnarray}
Here we have used the equation of state (\ref{ra}) in the last equality.
The AT line is determined by the vanishing point of the eigenvalue
$\tilde{\lambda}_+$,\cite{AGS2} or equivalently, $uv=1$, 
which has the following explicit form, 
\begin{equation}
   \alpha\beta^2 \ll \left(\langle \sigma_{\rho K}\sigma_{\rho L}\rangle
     -\langle \sigma_{\rho K} \rangle^2 \right)^2 \gg =\frac{q}{r}\ .
     \label{AT-1}
\end{equation}

It is necessary to evaluate the expectation values of spin variables
appearing in (\ref{AT-1}).  We define the quantity in the brackets 
$\ll\cdots\gg$ in (\ref{AT-1}) as $V$:
\begin{eqnarray}
    V&=&\ll \left(\langle \sigma_{\rho K}\sigma_{\rho L}\rangle
     -\langle \sigma_{\rho K} \rangle^2 \right)^2 \gg 
             \nonumber\\
     &=& \ll \langle \sigma_{\rho K}\sigma_{\rho L}\rangle
                  \langle \sigma_{\lambda K}\sigma_{\lambda L}\rangle\gg
              \nonumber\\
     & &   -2\ll \langle \sigma_{\rho K}\sigma_{\rho L}\rangle
              \langle\sigma_{\lambda K}\rangle
              \langle\sigma_{\sigma K}\rangle\gg
              \nonumber\\
     & &   +\ll \langle \sigma_{\rho K}\rangle 
             \langle \sigma_{\lambda K}\rangle
                 \langle \sigma_{\sigma K}\rangle 
                   \langle \sigma_{\kappa K}\rangle
                    \gg\ .
               \nonumber
\end{eqnarray}
According to the usual replica formalism,\cite{Binder-Young} this equation
is equivalent to the following expression in the limit $n\to 0$:

\begin{full}
\begin{eqnarray}
  V&=&\ll \sum_\sigma \left(
      \sigma_{\rho K}\sigma_{\rho L}\sigma_{\lambda K}\sigma_{\lambda L}
   -2\sigma_{\rho K}\sigma_{\rho L}\sigma_{\lambda K}\sigma_{\nu K}
   +  \sigma_{\rho K}\sigma_{\lambda K}\sigma_{\nu K}\sigma_{\kappa L}\right)
       \nonumber\\
 & &\times
   \exp \left[\frac{\beta}{M}{\mib m}\cdot{\mib \xi}
   \sum_{K\rho}\sigma_{\rho K}
    +B\sum_{K\rho}\sigma_{\rho K}\sigma_{\rho ,K+1}
    +\frac{\alpha\beta^2}{2M^2}r
      \sum_{KL\rho\sigma}\sigma_{\rho K}\sigma_{\sigma L}
       \right] \gg\ ,
       \label{sigma-Rep}
\end{eqnarray}
\end{full}

\noindent
where different suffixes $\rho, \lambda, \nu$ and $\kappa$
correspond to different replicas.

Let us consider the single-retrieval case $m_\mu =\delta_{\mu 1}\cdot m$.
The double summation over $K, L, \rho ,\sigma$ in (\ref{sigma-Rep})
can be decoupled using the Gaussian integral. 
In the limit $n\to 0$, we find
\begin{equation}
  V=\int {\rm D}z \left(\langle \sigma_K \sigma_L \rangle_z
         -\langle \sigma_K \rangle_z^2 \right)^2\ ,
         \label{V}
\end{equation}
where $\langle\cdots\rangle_z$ stands for the average 
with respect to the weight
\begin{equation}
  \exp \left[\frac{\beta m}{M}\sum_K \sigma_K+B\sum_K \sigma_K \sigma_{K+1}
    +\frac{\beta \sqrt{\alpha r}}{M}z\sum_K \sigma_K \right]\ .
\end{equation}
The averages appearing in the above equation can be calculated easily.
The single-spin average is given by 
\begin{eqnarray}
    \langle \sigma_K \rangle_z &=&\frac{{\rm Tr}\,\sigma_z
     {\rm e}^{\beta h\sigma_z +\beta\Delta \sigma_x}}
      {{\rm Tr}\, {\rm e}^{\beta h\sigma_z +\beta\Delta \sigma_x}}
        \nonumber\\
    &=&\frac{h}{\sqrt{h^2+\Delta^2}} \tanh \beta\sqrt{h^2+\Delta^2}\ ,
    \label{s-site}
\end{eqnarray}
with $h=m+\sqrt{\alpha r}z$, and the first equality holds in the limit 
$M\to\infty$.  The two-spin expectation value in (\ref{V}) is the 
correlation function of the one-dimensional Ising model in a uniform 
magnetic field, and thus can be calculated by the transfer-matrix
method. The result is, in the limit $M\to\infty$, 
\begin{equation}
  \langle \sigma_K \sigma_L \rangle_z =\frac{h^2}{v^2}
    +\frac{\Delta^2 \cosh\beta v(1-2y)}{v^2 \cosh\beta v}\ ,
  \label{correlation}
\end{equation}
with $v=(h^2+\Delta^2)^{1/2}$ and $y=(K-L)/M$.

The last expression (\ref{correlation}) of the correlation 
function has dependence on the Trotter numbers $K$ and $L$ 
through $y$, though we have calculated 
it under the static approximation which ignores such dependence.
This inconsistency is remedied if we average (\ref{correlation}) 
over the interval $0\le y\le 1$.  For example, the ground-state 
expression of the order parameter $S(=q)$ in the static 
approximation (\ref{eq-q}) can be obtained by the integral:
\begin{equation}
  \lim_{T\to 0}\int_0^1 {\rm d}y\langle \sigma_K \sigma_L \rangle_z 
 =\int{\rm D}z\frac{\left(m+\sqrt{\alpha r}z\right)^2}
                  {\left(m+\sqrt{\alpha r}z\right)^2 +\Delta^2}\ .
\end{equation}
This result seems quite natural in consideration of the definition 
(\ref{SK}) of the order parameter $S(KL)$.  Therefore, we replace 
the quantity in the brackets in (\ref{V}) by its average, 
\begin{eqnarray}
  \int_0^1 {\rm d}y\left(\langle \sigma_K \sigma_L \rangle_z
      -\langle \sigma_K \rangle_z^2 \right)
  &=&\frac{h^2}{v^2 \cosh^2 \beta v}+\frac{\Delta^2}{\beta v^3}
      \tanh \beta v
       \nonumber \\
  &\longrightarrow&\frac{T\Delta^2}{v^3}~~~~(T\to 0)\ .
       \nonumber
\end{eqnarray}
Equation (\ref{V}) now can be written as
\[
  V=T^2 \Delta^4 \int \frac{{\rm D}z}{v^3}\ ,
  \]
and the AT line is finally obtained from this relation and (\ref{AT-1}) as 
  \[
    q=\alpha r \Delta^4 \int\frac{{\rm D}z}
      {\left[\left(m+\sqrt{\alpha r}z\right)^2+\Delta^2\right]^3}\ .
   \]
  

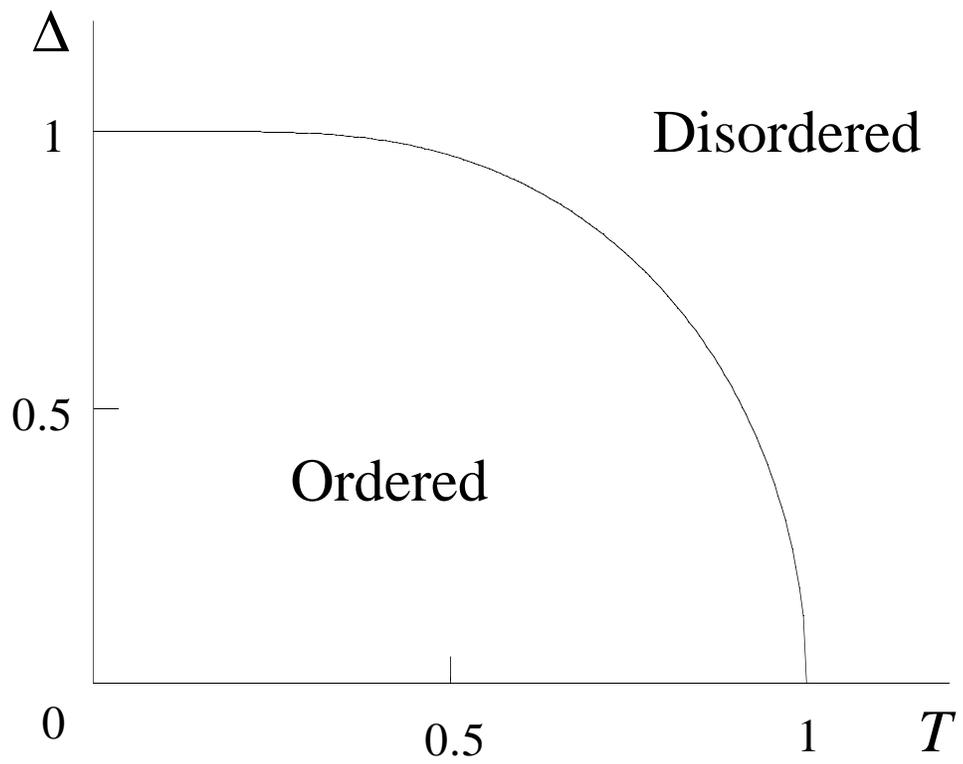
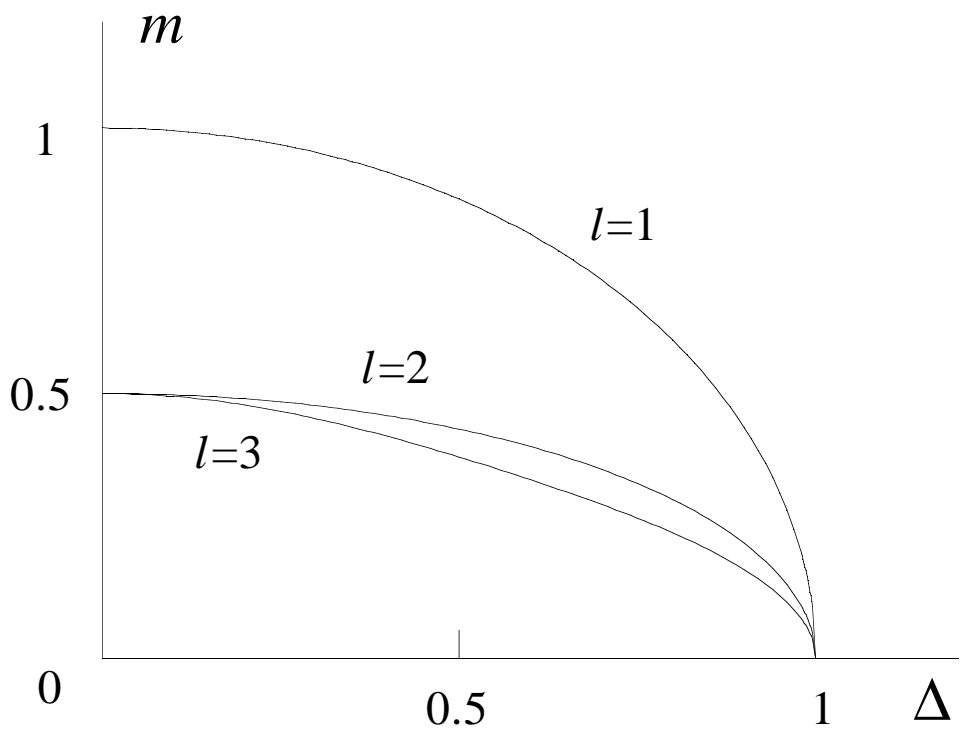
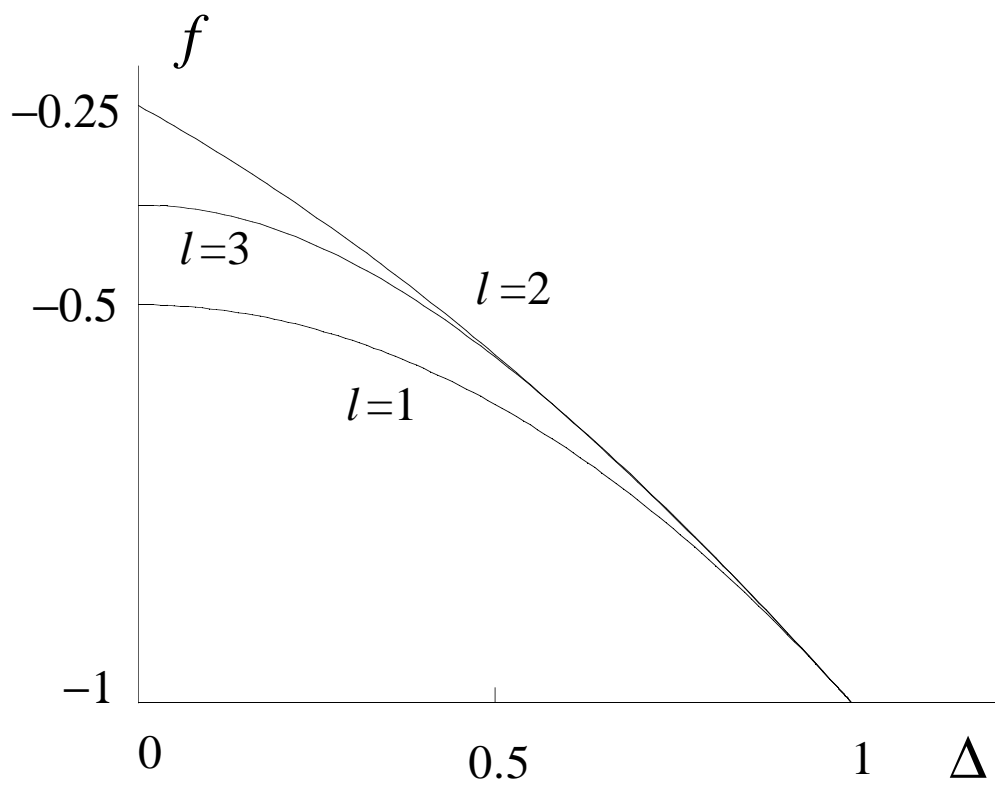
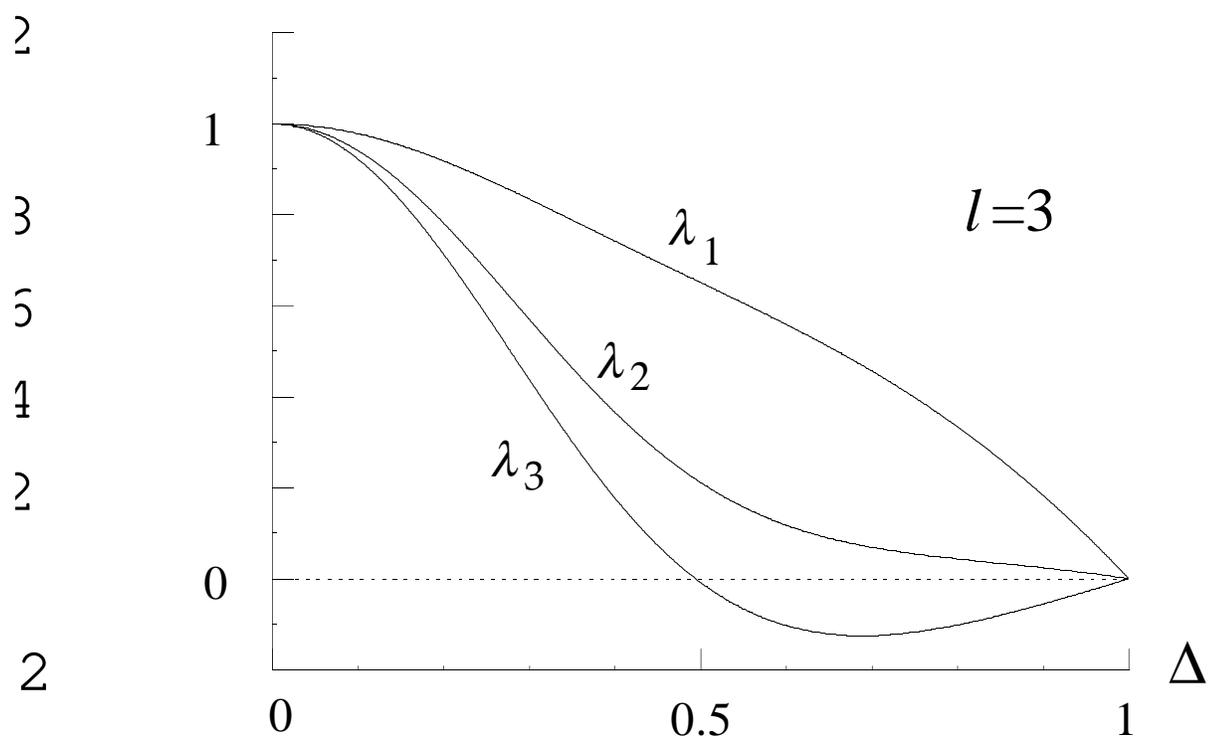
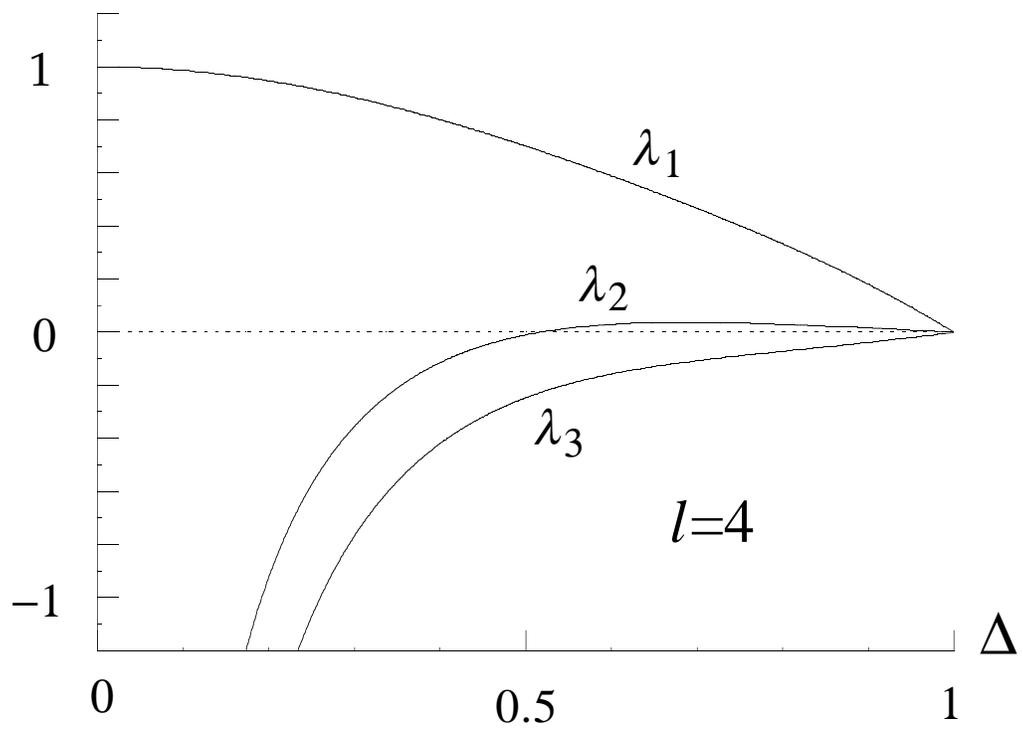
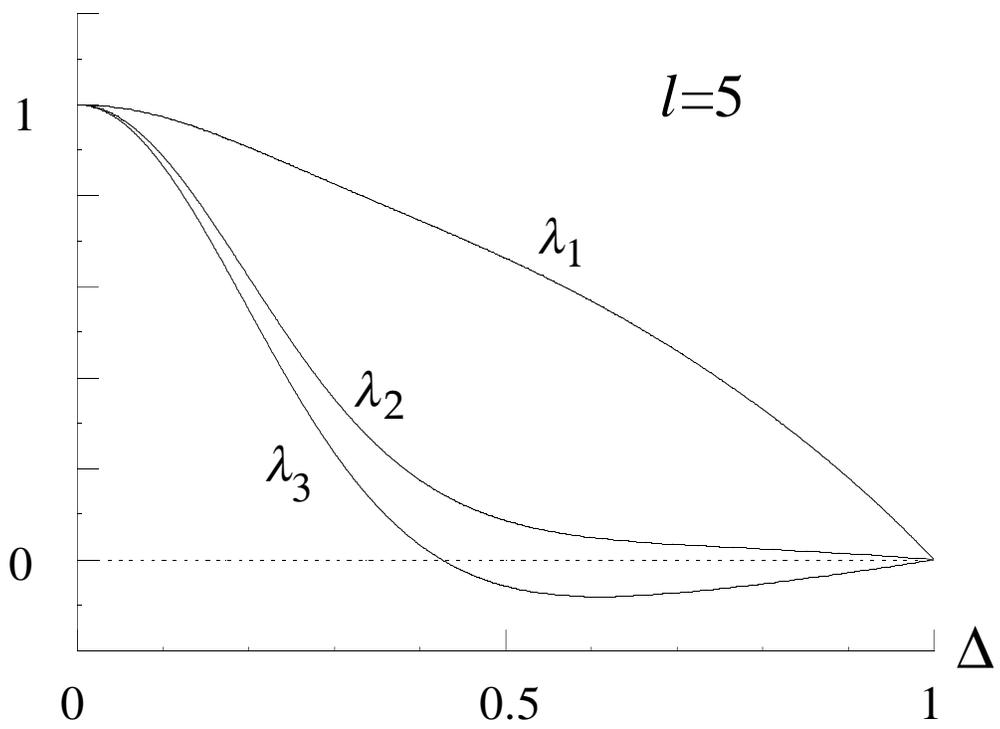
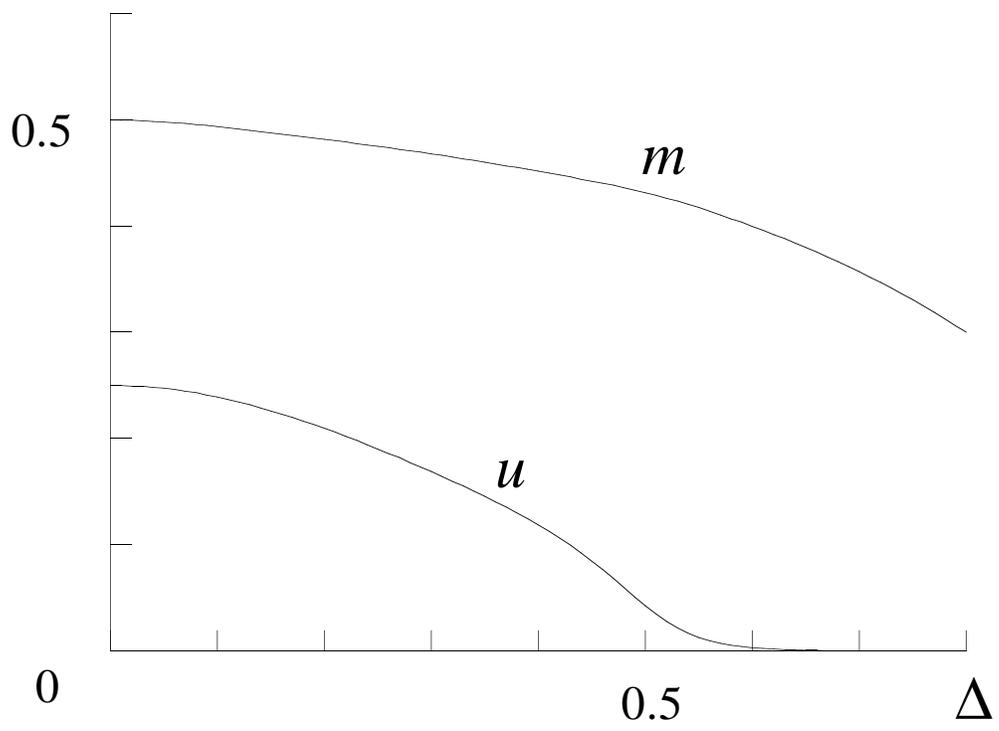
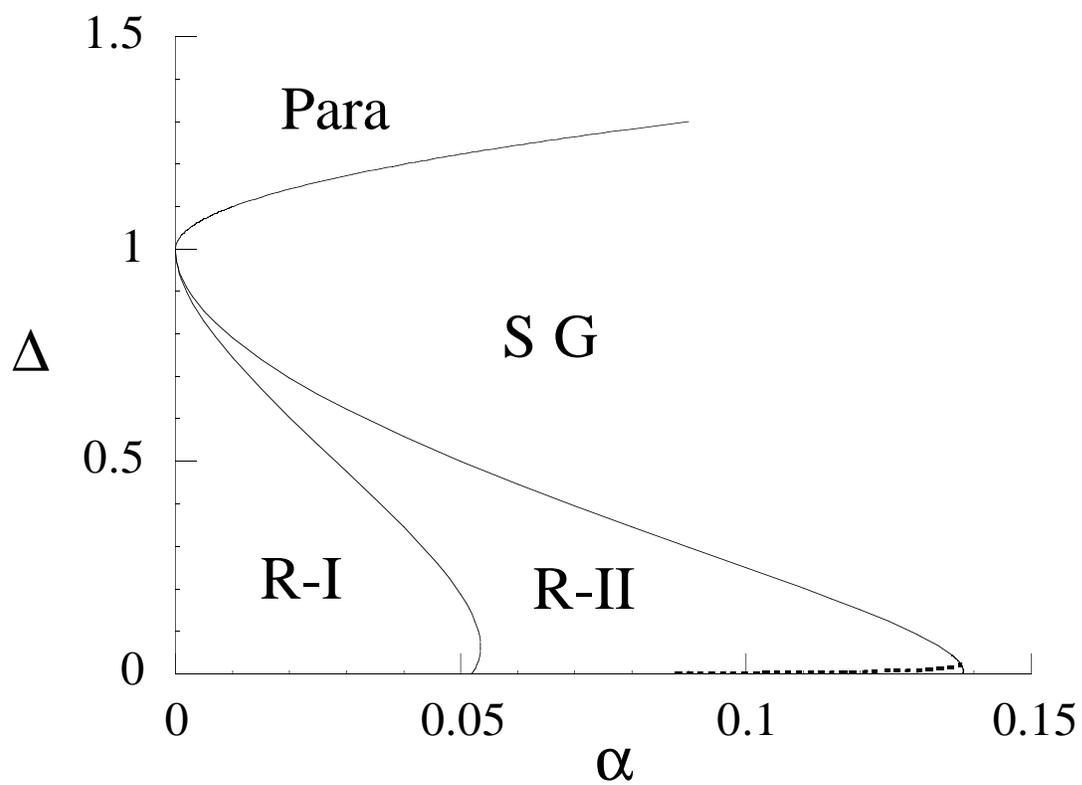
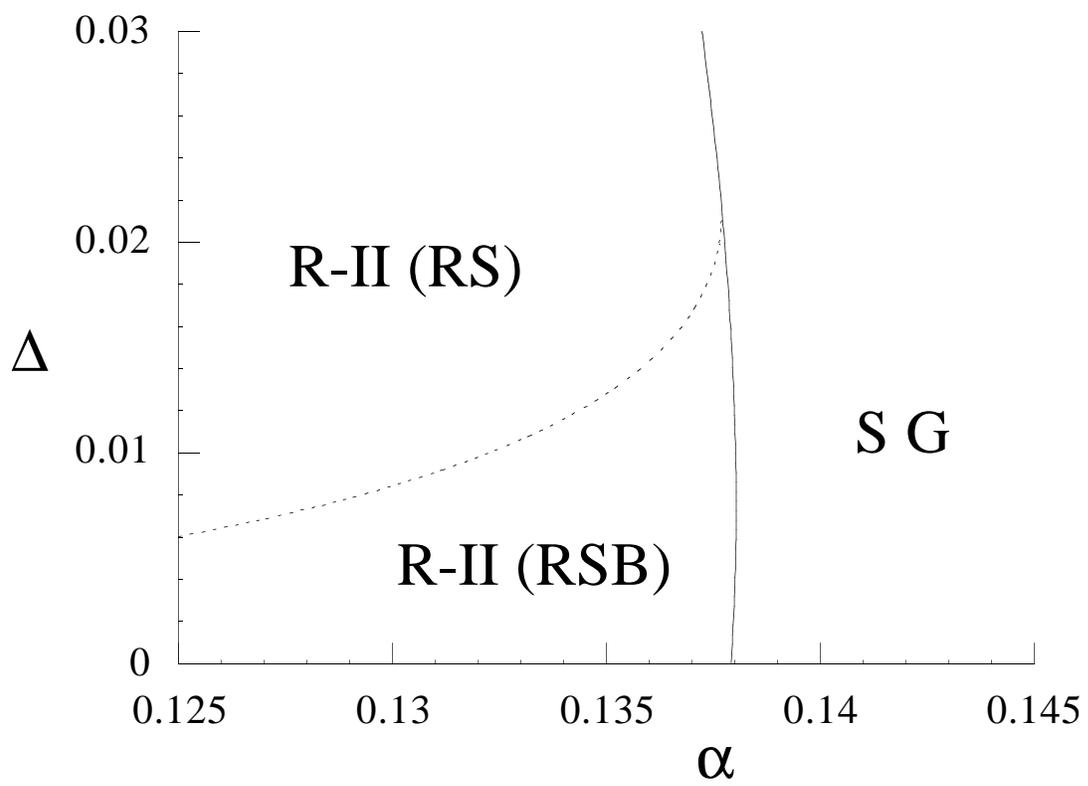
\end{document}